\documentclass[fleqn,usenatbib]{mnras}

\usepackage{newtxtext,newtxmath}

\usepackage[T1]{fontenc}
\usepackage{ae,aecompl}

\usepackage{graphicx}	
\usepackage{amsmath}	

\usepackage{multirow}
\usepackage[percent]{overpic}

\newcommand{\brvs}{Br\"unt-V\"ais\"al\"a}

\title[Day-Night Atmospheric Models of Exoplanets]{Coupled Day-Night Models of Exoplanetary Atmospheres}

\author[Gandhi and Jermyn]{
Siddharth Gandhi,$^{1,2}$\thanks{E-mail: Siddharth.Gandhi@warwick.ac.uk}
Adam S. Jermyn,$^{3}$\thanks{E-mail: adamjermyn@gmail.com}
\\
$^{1}$Department of Physics, University of Warwick, Coventry CV4 7AL, UK\\
$^{2}$Centre for Exoplanets and Habitability, University of Warwick, Gibbet Hill Road, Coventry CV4 7AL, UK \\
$^{3}$Center for Computational Astrophysics, Flatiron Institute, New York, NY 10010, USA
}

\date{Accepted XXX. Received YYY; in original form ZZZ}

\pubyear{2020}

\begin{document}
\label{firstpage}
\pagerange{\pageref{firstpage}--\pageref{lastpage}}
\maketitle

\begin{abstract}
We provide a new framework to model the day side and night side atmospheres of irradiated exoplanets using 1-D radiative transfer by incorporating a self-consistent heat flux carried by circulation currents (winds) between the two sides.
The advantages of our model are its physical motivation and computational efficiency, which allows for an exploration of a wide range of atmospheric parameters. 
We use this forward model to explore the day and night side atmosphere of WASP-76~b, an ultra-hot Jupiter which shows evidence for a thermal inversion and Fe condensation, and WASP-43~b, comparing our model against high precision phase curves and general circulation models. We are able to closely match the observations as well as prior theoretical predictions for both of these planets with our model.
We also model a range of hot Jupiters with equilibrium temperatures between 1000-3000~K and reproduce the observed trend that the day-night temperature contrast increases with equilibrium temperature up to $\sim$2500~K beyond which the dissociation of H$_2$ becomes significant and the relative temperature difference declines.
\end{abstract}




\section{Introduction}

Observations of exoplanet atmospheres have advanced tremendously in recent years.
Numerous phase curves for hot Jupiters from Spitzer observations have provided thermal emission spectra throughout the planetary orbit \citep[e.g.][]{knutson2007, knutson2009, knutson2009b, knutson2012, stevenson2017} resulting in constraints on atmospheric properties as a function of planetary longitude \citep[e.g.][]{cowan2008}.
Likewise, HST WFC3 (Hubble Space Telescope Wide Field Camera 3) has also produced phase resolved spectra for a number of hot and ultra-hot Jupiters \citep{stevenson2014, kreidberg2018, arcangeli2019}, and advances in high resolution spectroscopy have enabled constraints on the winds on hot Jupiters \citep{snellen2010, brogi2016} and chemical variation between their day and night sides \citep{ehrenreich2020}. 

Three-dimensional general circulation models (GCMs) have been crucial to interpreting these observations~\citep[e.g.][]{showman2008, showman2009, rauscher2010, dobbs-dixon2013, mayne2014, kataria2016, flowers2019, drummond2020}. GCMs simulate the dynamics of a planetary atmosphere in three dimensions, providing unparalleled levels of spatial and dynamical detail and often incorporating important processes such as chemical reactions and radiative heat transport \citep[e.g.][]{cooper2006, showman2011, rauscher2013, wordsworth2015, hammond2018}.

Because of their physical detail GCMs are often computationally expensive, and so 1-D and 2-D models have attracted interest as fast ways to simulate day side and night side atmosphere of exoplanets. \citet{burrows2008} introduced a combined day and night side 1-D model which transfers energy between the irradiated day to the unirradiated night side. \citet{koll2016} implemented a model for rocky planets which also solves for the resulting thermal equilibrium, and computed wind speeds by balancing dissipation against the work done by a heat engine. \citet{tremblin2017} used 2-D models to explore the inflated radii of hot Jupiters.
A key advantage of such models is that they allow for broad explorations of parameter space to interpret e.g. recent phase resolved HST observations \citep{stevenson2014, kreidberg2018, arcangeli2019} and high resolution spectroscopic measurements of winds and chemical variability \citep{snellen2010, brogi2016, ehrenreich2020}.
Such models can be used to determine the most physically interesting properties and parts of parameter space to explore in more detail with GCMs. In addition, simultaneous retrievals of hot Jupiter phase curves have also recently been carried out assuming more realistic parameter variation over the traditional 1-D approaches \citep{irwin2020}.

The aim of the present work is to introduce a self-consistent wind model that calculates both the day and night side temperature profiles of the atmosphere in thermal equilibrium. Our wind model is similar to that used in \citet{koll2016} in that we employ an energy balance argument, but we provide a more detailed prescription for energy dissipation which is specialised for use in gaseous planets. We transfer the energy between the day and night side by adapting the prescription in \citet{burrows2008} for the thermal equilibrium equations.

We model both the night side and day side self-consistently by including a wind heat flux between the two sides using the GENESIS atmospheric model \citep{gandhi2017}. We first compute the wind speed in Section~\ref{sec:theory} by balancing energy input from the day-night temperature difference against dissipation, modelled using turbulent scaling relations. From the resulting wind speed we then determine a heat flux.
This approach enables fast and efficient modelling of the day and night side while incorporating recent insights into turbulent dissipation processes~\citep[e.g.][]{0004-637X-837-2-133}.

In Section~\ref{sec: GENESIS modifying} we incorporate the heat flux into the GENESIS 1D radiative-convective equilibrium code, and we then validate our results against the work by~\citet{komacek2016} in Section~\ref{sec:validation}.

In Section~\ref{sec:results} we use our forward modelling framework to explore a number of properties of exoplanetary atmospheres. We first model the day and night side of WASP-76~b in radiative-convective and thermochemical equilibrium, an ultra-hot Jupiter which has shows evidence of a thermal inversion and Fe condensation \citep{fu2020, ehrenreich2020}. We model the thermal inversion on the day side with TiO and we find that the night side is cool enough for Fe to condense in the photosphere. 
We also model WASP-43~b in radiative-convective equilibrium and compare our results to GCMs \citep{kataria2015} and to the retrieved day and night side temperature profiles from phase curve observations \citep{stevenson2014, stevenson2017}. We find that our wind model shows good agreement to the temperature profile for pressures $\lesssim$0.1~bar, above which the data are not constraining.

Finally, we study the day-night temperature contrast as a function of equilibrium temperature to see how energy redistribution varies with irradiation. We model hot Jupiters with a wide range of equilibrium temperatures between 1000-3000~K and compare these against previous observations \citep[e.g][]{knutson2012, wong2016, zhang2018, kreidberg2018} and theory \citep{komacek2016, keating2019}.
We find good agreement, with the temperature contrast increasing with equilibrium temperature until H$_2$ dissociation becomes important, beyond $\sim$2500~K, at which point it begins to decline~\citep{bell2018, komacek2018}.

\section{Theory}\label{sec:theory}

\begin{figure*}
\begin{overpic}[width=\textwidth]{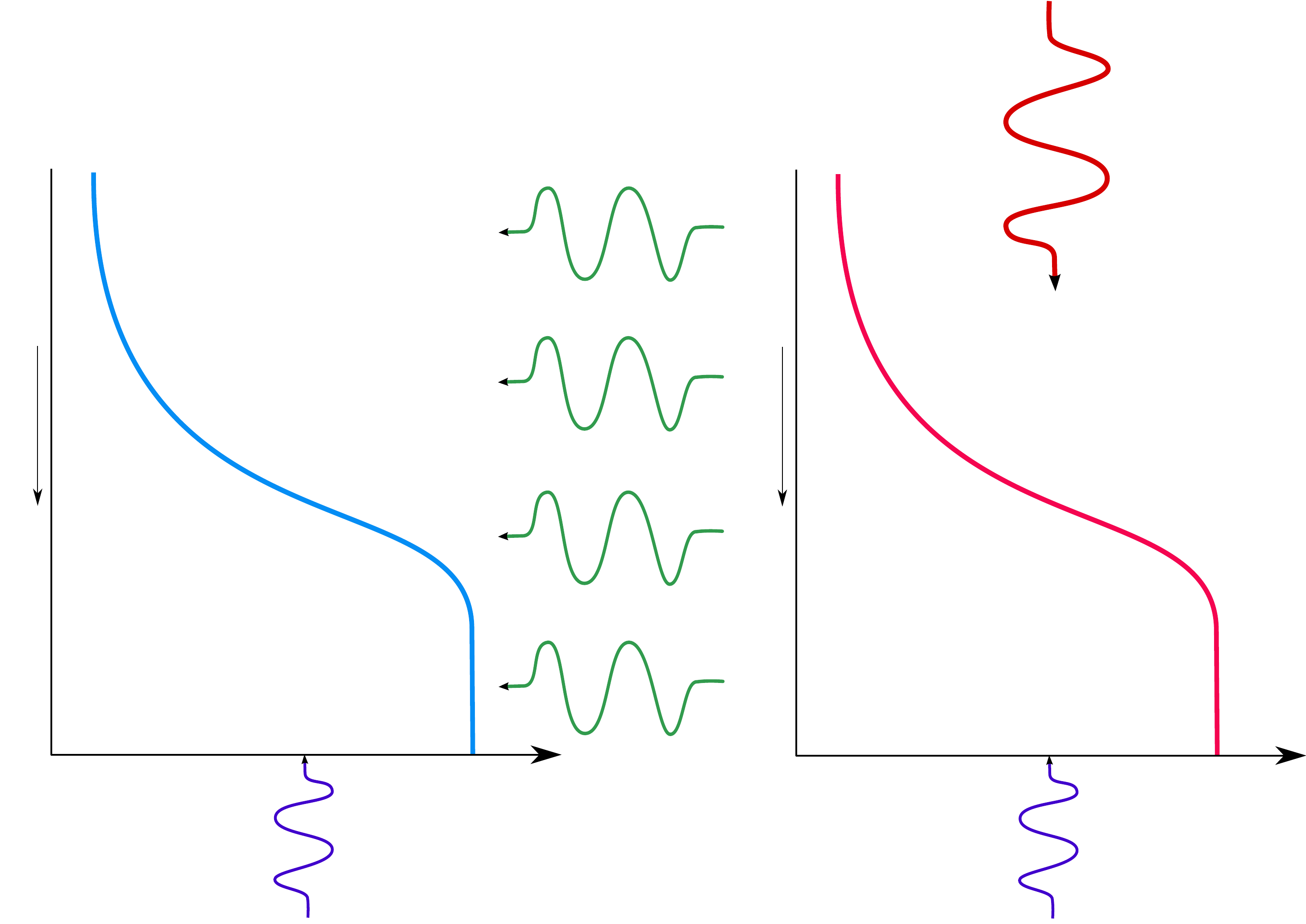}
 \put (40,10) {\LARGE $\mathrm{T}$}
 \put (97,10) {\LARGE $\mathrm{T}$}
 \put (25,1) {\huge $\mathrm{F_{int}}$}
 \put (82,1) {\huge $\mathrm{F_{int}}$}
 \put (43,34.5) {\huge $\mathrm{F_{wind}}$}
 \put (74,66) {\Huge $\mathrm{F_{ext}}$}
 \put (72,23) {\Huge $\mathrm{day \,side}$}
 \put (14,23) {\Huge $\mathrm{night \,side}$}
 \put (0,35) {\rotatebox{90}{\LARGE $\mathrm{log(P)}$}}
 \put (57,35) {\rotatebox{90}{\LARGE $\mathrm{log(P)}$}}
\end{overpic}
\caption{Schematic of the pressure-temperature profile and thermal fluxes in the day side and night side GENESIS model. Incident stellar flux from the top of the day side is absorbed by the day side atmosphere. This heat transfers through the thermal gradient to the night side via a depth dependent wind flux. The internal heat flux from the planetary core is assumed to be the same for both sides of the atmosphere. Section~\ref{sec:theory} discusses the theory and the numerical implementation is summarised in Section~\ref{sec: GENESIS modifying}.}
\label{fig:sketch}
\end{figure*}
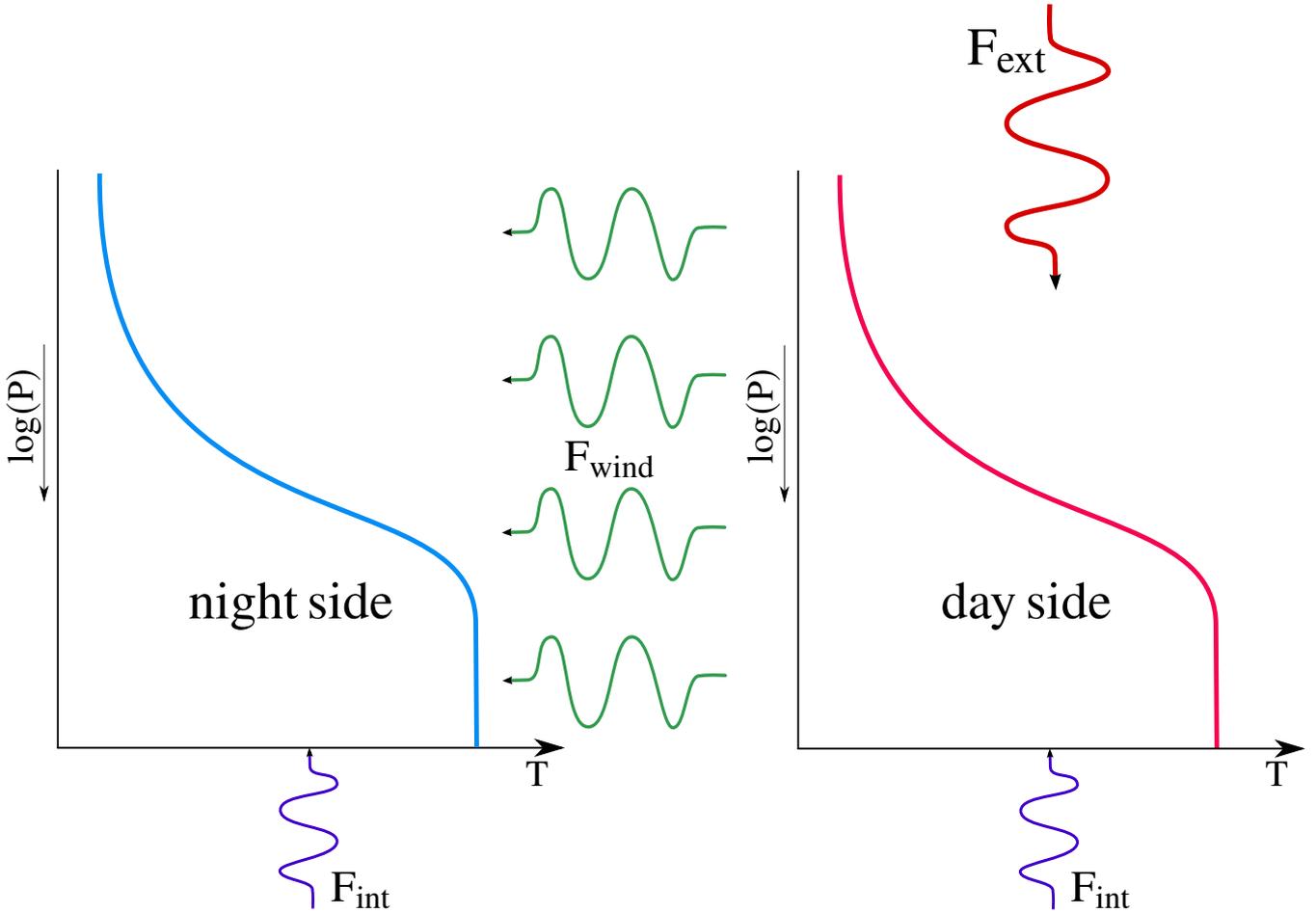

We now develop our wind model. We begin by outlining our key assumptions in Section~\ref{sec:assumptions}. We then compute the kinetic energy budget for the system, balancing dissipation against work done by a day-night temperature difference in Section~\ref{sec:power}. In Section~\ref{sec:radconv} we calculate the dissipation rates in radiative and convective zones. We then estimate the characteristic length-scales of the flow in Section~\ref{sec:scales} and put it all together in Section~\ref{sec:summary} to obtain the wind speed in a variety of different circumstances.
We then compute the heat flux carried by winds in Section~\ref{sec:flux}, and correct this for radiative losses in Section~\ref{sec:efficiency}.

\subsection{Assumptions}
\label{sec:assumptions}

In our analysis we make several simplifying assumptions.
First, we treat the atmosphere as being composed of a day side and a night side, each of which has its own pressure-temperature profile as shown in Figure~\ref{fig:sketch}.
That is, the day side is characterised by $T_{\rm day}(P)$ while the night side is characterised by $T_{\rm night}(P)$~\citep{doi:10.1093/mnras/stx831}.
The two sides interact only via a wind which transfers heat between the two.
This amounts to an expansion in spherical harmonics centred on the subsolar point, keeping keeping the modes with $(l,m)=(0,0)$ and $(l,m)=(1,0)$.
The amplitude of the former is
\begin{align}
	A_{00}(P) = T_{\rm avg} \equiv \frac{1}{2}\left(T_{\rm day}(P) + T_{\rm night}(P)\right),
    \label{eq:tavg}
\end{align}
while that of the latter is
\begin{align}
	A_{20}(P) = \frac{1}{2}\Delta T(P) = \frac{1}{2}\left(T_{\rm day}(P) - T_{\rm night}(P)\right).
    \label{eq:dt}
\end{align}
This expansion enables us to solve the equations of radiative transfer on each side separately, with a sink term on the day side and an equal source term on the night side.
The job of the wind model is to provide the magnitude of this heat transfer given the pressure-temperature profile on either side. We have assumed here that the day side receives the full incident radiation and the night side receives none, but our model could be extended in the future to model more longitudes, thereby allowing for a 1.5-D approach considering incident radiation that is reduced by the cosine of the longitude.

Next, we take the wind to flow primarily along surfaces of constant density (i.e. isochors), so that the flow is two-dimensional.
This enables us to consider the flow at different densities independently and so allows us to use one-dimensional radiative transfer models to calculate the heat balance on each of the day and night sides.
We justify this approximation in Appendix~\ref{sec:2D}.
Because pressure is a more natural coordinate for our radiative transfer methods we shall actually identify points of equal pressure between the two sides rather than points of equal density, but the two are similar enough that this does not entail significant error.

A further approximation is required to handle the possibility of turbulence.
Turbulent systems exhibit fluctuations in the velocity and other fields.
We average over these, so that $\boldsymbol{u}$ refers to an average velocity over long time-scales.
This procedure results in a turbulent contribution to the effective viscosity and stress, which we incorporate into our equations.

Finally, following~\citet{https://doi.org/10.7907/z90z716m} and~\citet{koll2016} we treat the wind as being in energetic steady state, so that the input of energy from the temperature gradient balances losses due to viscous effects, including turbulent viscosity as appropriate, and thermal diffusion.
This allows us to determine the speed $u$ of the flow, which we do in Section~\ref{sec:power}.

At various points in this analysis we shall make claims which are verifiable only at the end once the answer is known.
We verify these in Appendix~\ref{appen:verify}.

\subsection{Power Balance}
\label{sec:power}

We now aim to compute the speed $u$ of the flow by balancing the kinetic energy budget the system, following the reasoning of~\citet{https://doi.org/10.7907/z90z716m}.
We do this rather than computing a force balance because there are conservative forces, such as the Coriolis effect, which do not contribute to the energy budget of the flow.
Likewise the acceleration owing to gravity and the behaviour of the pressure gradient are more readily analysed in this way.

To begin note that the kinetic energy density of the fluid is
\begin{align}
	K \equiv \frac{1}{2}\rho u^2,
\end{align}
This obeys the equation
\begin{align}
	\frac{D K}{D t} = \rho \boldsymbol{u}\cdot\frac{D\boldsymbol{u}}{D t} + \frac{1}{2}u^2\frac{D\rho}{Dt},
	\label{eq:dk}
\end{align}
where
\begin{align}
	\frac{D}{Dt} \equiv \frac{\partial}{\partial t} + \boldsymbol{u}\cdot\nabla
\end{align}
denotes the material derivative.
The second term in equation~\eqref{eq:dk} may be written as
\begin{align}
	\frac{D\rho}{Dt} = -\rho\nabla\cdot\boldsymbol{u}.
    \label{eq:drho}
\end{align}
To expand the first term we make use of the Navier-Stokes equation, which yields
\begin{align}
	\frac{D \boldsymbol{u}}{D t} + 2\boldsymbol{\Omega}\times\boldsymbol{u} = \boldsymbol{g} - \frac{1}{\rho}\nabla P + \boldsymbol{V}(\boldsymbol{u}) + \boldsymbol{\Lambda},
    \label{eq:navierstokes}
\end{align}
where $\boldsymbol{g}$ is the effective gravitational field accounting for the centrifugal acceleration.
The forcing owing to turbulence is given by $\boldsymbol{\Lambda}$, which is independent of the velocity and present only in convection zones~\citep{1993A&A...276...96K}.
The viscosity term $\boldsymbol{V}(\boldsymbol{u})$ incorporates both microscopic and turbulent components and vanishes when the velocity field vanishes\footnote{This may be recast as an effective drag time-scale along the lines of~\citet{komacek2016} via $\tau_{\rm drag} = \frac{u^2}{\boldsymbol{u}.\boldsymbol{V}(\boldsymbol{u})}.$}
Note that because the microscopic viscosity is vanishingly small we are only ever be concerned with the turbulent part, which we shall describe in more detail later.

We have neglected the magnetic field because the ionization fraction in the atmospheres of planets is typically low, though in the hottest planets such effects may become important.
More specifically, the magnetic field contributes to equation~\eqref{eq:dk} an amount of order $f B^2 u (4\pi l)^{-1}$, where $f$ is the ionization fraction, $B$ is the magnetic field strength and $l$ is the characteristic length-scale over which the field varies.
By contrast we shall see that the temperature gradient supplies power of order $\rho u g r^{-1} T_{\rm avg}^{-1}\Delta T$. 
The ratio of the latter to the former is $4\pi \rho l g T_{\rm avg}^{-1}\Delta T (f B^2)^{-1}$.
Suppose $l \ga h$, $h \approx 10^7\mathrm{cm}$, $g \approx 10^{3}\mathrm{cm\,s^{-2}}$, and take the generous bound $B \la 10^3 \mathrm{G}$.
The thermal term is then a factor of at least $10^5 (\rho/\mathrm{g\,cm^{-3}}) f^{-1} T_{\rm avg}^{-1}\Delta T$ stronger than the magnetic term.
Except at very low densities high in the atmosphere this is comfortably greater than unity, so we expect this to be a good approximation.

Inserting equations~\eqref{eq:drho} and~\eqref{eq:navierstokes} into equation~\eqref{eq:dk} we obtain
\begin{align}
	\frac{D K}{D t} = \rho \boldsymbol{u}\cdot\left[\boldsymbol{g} -2\boldsymbol{\Omega}\times\boldsymbol{u}- \frac{1}{\rho}\nabla P + \boldsymbol{V}(\boldsymbol{u}) + \boldsymbol{\Lambda}\right] - \frac{1}{2} \rho u^2 \nabla\cdot\boldsymbol{u}.
	\label{eq:yes_coriolis}
\end{align}
Because the cross product of one vector with another is orthogonal to both, the Coriolis term vanishes and
\begin{align}
	\frac{D K}{D t} = \rho \boldsymbol{u}\cdot\left[\boldsymbol{g} - \frac{1}{\rho}\nabla P + \boldsymbol{V}(\boldsymbol{u}) + \boldsymbol{\Lambda}\right] - \frac{1}{2} \rho u^2 \nabla\cdot\boldsymbol{u}.
	\label{eq:no_coriolis}
\end{align}
In order for the system to be in steady state the total kinetic energy must not change, so
\begin{align}
	\int \frac{D K(\boldsymbol{r},t)}{D t} d^3\boldsymbol{r} = 0,
\end{align}
where the integral is over the entire system.
Hence
\begin{align}
	0&=\int \rho(\boldsymbol{r}) \boldsymbol{u}(\boldsymbol{r})\cdot\left[\boldsymbol{g}(\boldsymbol{r}) - \frac{1}{\rho(\boldsymbol{r})}\nabla P(\boldsymbol{r}) + \boldsymbol{V}(\boldsymbol{u}) + \boldsymbol{\Lambda}(\boldsymbol{r})\right]d^3\boldsymbol{r}\nonumber\\
    &- \frac{1}{2} \int\rho(\boldsymbol{r}) u^2(\boldsymbol{r}) \nabla\cdot\boldsymbol{u}(\boldsymbol{r}) d^3\boldsymbol{r}.
\label{eq:balance}
\end{align}

We now demonstrate a helpful result, which is that
\begin{align}
	\int \rho(\boldsymbol{r}) \boldsymbol{u}(\boldsymbol{r}) \cdot \nabla f(\boldsymbol{r}) d^3\boldsymbol{r} = 0
    \label{eq:fres}
\end{align}
for any differentiable function $f(\boldsymbol{r})$.
This is because the divergence theorem implies that
\begin{align}
	\int \rho(\boldsymbol{r}) \boldsymbol{u}(\boldsymbol{r}) \cdot \nabla f(\boldsymbol{r}) d^3\boldsymbol{r} &= \int_{\mathcal{S}} \rho(\boldsymbol{r}) f(\boldsymbol{r}) \boldsymbol{u}(\boldsymbol{r}) \cdot d\mathcal{S}\nonumber\\
    &- \int\nabla\cdot( \rho(\boldsymbol{r})\boldsymbol{u}(\boldsymbol{r})) f(\boldsymbol{r}) d^3\boldsymbol{r},
\end{align}
where $\mathcal{S}$ is a closed surface containing the integration volume and $d\mathcal{S}$ is the differential surface element.
Because the integration volume is the entire planet the surface $\mathcal{S}$ lies outside the planet where $\rho$ vanishes.
Hence the first term vanishes.
The second term vanishes by mass conservation in steady state, so the result holds.

Because $\boldsymbol{g}$ is the gradient of a potential equation~\eqref{eq:fres} implies that its contribution to equation~\eqref{eq:balance} is zero, so
\begin{align}
	0&=\int \rho(\boldsymbol{r}) \boldsymbol{u}(\boldsymbol{r})\cdot\left[ - \frac{1}{\rho(\boldsymbol{r})}\nabla P(\boldsymbol{r}) + \boldsymbol{V}(\boldsymbol{u},\boldsymbol{r}) + \boldsymbol{\Lambda}(\boldsymbol{r})\right] d^3\boldsymbol{r}\nonumber\\
    &- \frac{1}{2}\int \rho(\boldsymbol{r}) u^2(\boldsymbol{r}) \nabla\cdot\boldsymbol{u}(\boldsymbol{r}) d^3\boldsymbol{r}.
\label{eq:balance2}
\end{align}
Along similar lines note that if $p$ is purely a function of $\rho$ then there exists a function $q(\rho)$
\begin{align}
	q(\rho) \equiv \int_0^{\rho} \frac{\nabla P(\rho')}{\rho'} d\rho',
\label{eq:q0}
\end{align}
such that
\begin{align}
\nabla q = \frac{1}{\rho}\nabla P.
\end{align}
In this case the contribution of the pressure gradient to equation~\eqref{eq:balance2} vanishes by equation~\eqref{eq:fres}.
Of course this is a somewhat unusual limit.
More realistically note that, neglecting variations in composition, the equation of state allows us to write
\begin{align}
	\nabla P = \left.\frac{\partial p}{\partial T}\right|_{\rho}\nabla T + \left.\frac{\partial p}{\partial \rho}\right|_{T}\nabla \rho.
    \label{eq:eos}
\end{align}
In analogue to equation~\eqref{eq:q0} we define
\begin{align}
	q(\boldsymbol{r}) \equiv \int_0^{\boldsymbol{r}} \frac{\nabla \rho(\boldsymbol{r}')\cdot\nabla P(\boldsymbol{r}')}{|\nabla\rho(\boldsymbol{r}')|\rho(\boldsymbol{r}')} d\boldsymbol{r}',
\end{align}
where the integral proceeds along a path following the density gradient.
With this we see that
\begin{align}
	\nabla q = \nabla \rho\frac{\nabla \rho\cdot\nabla P}{|\nabla\rho|^2\rho}.
\end{align}
Inserting equation~\eqref{eq:eos} we find
\begin{align}
	\nabla q = \frac{1}{\rho}\nabla \rho\left(\left.\frac{\partial p}{\partial \rho}\right|_{T} + \left.\frac{\partial p}{\partial T}\right|_{\rho}\frac{\nabla \rho\cdot\nabla T}{|\nabla\rho|^2}\right),
\end{align}
so
\begin{align}
	\frac{1}{\rho} \nabla P - \nabla q = \frac{1}{\rho}\left.\frac{\partial p}{\partial T}\right|_{\rho}\left(1-\mathbb{P}_{\nabla \rho}\right) \nabla T,
\end{align}
where $\mathbb{P}_{\nabla \rho}$ is the operator which projects a vector along $\nabla \rho$.
Because this operator is linear we may also write this as
\begin{align}
	\frac{1}{\rho} \nabla P - \nabla q = \frac{P}{\rho}\left.\frac{\partial \ln p}{\partial \ln T}\right|_{\rho}\left(1-\mathbb{P}_{\nabla \rho}\right) \nabla \ln T.
\end{align}
For an ideal gas the logarithmic derivative is $1$, so
\begin{align}
	\frac{1}{\rho} \nabla P - \nabla q = \frac{P}{\rho}\left(1-\mathbb{P}_{\nabla \rho}\right) \nabla \ln T.
\label{eq:pert}
\end{align}
Hence by equation~\eqref{eq:fres} we find
\begin{align}
	0&=\int \rho(\boldsymbol{r}) \boldsymbol{u}(\boldsymbol{r})\cdot\left[ -\frac{P}{\rho}\left(1-\mathbb{P}_{\nabla \rho}\right) \nabla \ln T + \boldsymbol{V}(\boldsymbol{u},\boldsymbol{r}) + \boldsymbol{\Lambda}(\boldsymbol{r})\right] d^3\boldsymbol{r}\nonumber\\
    &- \frac{1}{2}\int \rho(\boldsymbol{r}) u^2(\boldsymbol{r}) \nabla\cdot\boldsymbol{u}(\boldsymbol{r}) d^3\boldsymbol{r}.
\label{eq:balance3}
\end{align}
This form is more useful than equation~\eqref{eq:balance2} because even when the system is spherically symmetric $\nabla P$ does not vanish, whereas in this form it is clear that that term contributes nothing in the symmetric limit.

With equation~\eqref{eq:steadycont} we may rewrite the final term of equation~\eqref{eq:balance3} and find
\begin{align}
	0&=\int \rho(\boldsymbol{r}) \boldsymbol{u}(\boldsymbol{r})\cdot\left[ -\frac{P}{\rho}\left(1-\mathbb{P}_{\nabla \rho}\right) \nabla \ln T + \boldsymbol{V}(\boldsymbol{u},\boldsymbol{r}) + \boldsymbol{\Lambda}(\boldsymbol{r})\right.\nonumber\\
    &\left.+ \frac{1}{2} u^2(\boldsymbol{r})\nabla \ln \rho(\boldsymbol{r})\right] d^3\boldsymbol{r}.
\label{eq:balance4}
\end{align}
This makes it clear that all terms in the equation are determined by their projection along $\boldsymbol{u}$.
We claim now and shall verify later in Appendix~\ref{appen:verify} that the final term in this equation is always small relative to the other components.
For now we drop this term and find
\begin{align}
	0&=\int \rho(\boldsymbol{r}) \boldsymbol{u}(\boldsymbol{r})\cdot\left[ -\frac{P}{\rho}\left(1-\mathbb{P}_{\nabla \rho}\right) \nabla \ln T + \boldsymbol{V}(\boldsymbol{u},\boldsymbol{r}) + \boldsymbol{\Lambda}(\boldsymbol{r})\right] d^3\boldsymbol{r}.
\label{eq:balance5}
\end{align}

We now examine equation~\eqref{eq:balance5} from an order of magnitude perspective.
The flow is confined to isochors so we may consider just the integral over one such surface.
There is a balance then between three terms, namely the temperature gradient, the viscosity and the turbulent forcing.
None of these are forced to be perpendicular to the flow, indeed the projection operator maps the temperature gradient into the flow plane and the turbulent terms are closely related to the velocity.
As such we expect the inner product in equation~\eqref{eq:balance5} to produce a factor of order unity.
Hence we approximate that equation by positing a balance between the three terms in brackets.

Because the temperature gradient is set externally while the turbulent forcing is set by the unperturbed equilibrium structure of the planet the two cannot be tuned to cancel each other, so we expect the power input to be at least as great at the larger of these.
Because only the viscous term explicitly depends on the velocity we write
\begin{align}
	|\boldsymbol{V}(\boldsymbol{u})| \approx \max\left(|\boldsymbol{\Lambda}|,\left|\frac{P}{\rho}\left(1-\mathbb{P}_{\nabla \rho}\right)\nabla \ln T\right|\right),
\end{align}
which serves to set the velocity.
For simplicity we have taken the maximum of the drivers, though other prescriptions such as adding them would also be acceptable at this level of accuracy.

Recalling equations~\eqref{eq:tavg} and~\eqref{eq:dt} we find
\begin{align}
\left(1-\mathbb{P}_{\nabla \rho}\right)\nabla \ln T \approx \frac{\Delta T}{T_{\rm avg}\pi R},
\end{align}
so
\begin{align}
	|\boldsymbol{V}(\boldsymbol{u})| \approx \max\left(|\boldsymbol{\Lambda}|,\left|\frac{P}{\rho}\frac{\Delta T}{T_{\rm avg}\pi R}\right|\right).
\label{eq:powerbalance}
\end{align}

As a further simplification we treat the viscous term as depending only on the direction of the shear and not on the direction of the flow.
In particular, following~\citet{1992A&A...265..115Z}, we consider a horizontal viscosity which couples to shears in the $\theta$ and $\phi$ directions and a vertical viscosity which couples to those in the radial direction, so that
\begin{align}
	|\boldsymbol{V}(\boldsymbol{u})| = \nu_h |\nabla_h^2 \boldsymbol{u}| + \nu_v |\nabla_v^2 \boldsymbol{u}|,
\end{align}
where $\nabla_h$ denotes the gradient in the horizontal directions, $\nabla_v$ denotes that in the vertical direction, $\nu_h$ is the horizontal viscosity and $\nu_v$ is the vertical viscosity.
To further simplify this expression we define $l_h$ and $l_v$ respectively as the characteristic length-scales on which the velocity changes in the horizontal and vertical directions.
Hence
\begin{align}
	|\boldsymbol{V}(\boldsymbol{u})| \approx \frac{\nu_h}{l_h^2} u + \frac{\nu_v}{l_v^2}u,
\end{align}
so that equation~\eqref{eq:powerbalance} reads
\begin{align}
	u\left(\frac{\nu_h}{l_h^2} + \frac{\nu_v}{l_v^2}\right) \approx \max\left(|\boldsymbol{\Lambda}|,\left|\frac{P}{\rho}\frac{\Delta T}{T_{\rm avg}\pi R}\right|\right).
\label{eq:powerbalanc2}
\end{align}

\subsection{Dissipation}
\label{sec:radconv}

In order to use equation~\eqref{eq:powerbalanc2} we must estimate the turbulent viscosities $\nu_h$ and $\nu_v$.
Because dissipation is very different in radiative and convective regions we analyze these cases separately.
Note that in the convective case we also consider the turbulent forcing $\boldsymbol{\Lambda}$, because the scale of that term is closely related to the scale of the convective turbulent viscosity.

\subsubsection{Radiative Zones}
\label{sec:radvisc}

In radiative zones the horizontal viscosity takes the form~\citep{1992A&A...265..115Z}
\begin{align}
	\nu_h \approx l_h u.
\label{eq:nuh}
\end{align}
For the vertical viscosity we use the doubly-diffusive prescription of~\citet{0004-637X-837-2-133} and neglect the correction owing to non-zero microscopic viscosity.
Hence
\begin{align}
	\nu_v \approx \frac{C}{1+aJ {\rm Pe})^{-1}} \frac{\alpha}{J},
    \label{eq:nuv0}
\end{align}
where
\begin{align}
	J \equiv \frac{N^2 l_v^2}{u^2 }
\end{align}
is the Richardson number,
\begin{align}
	{\rm Pe} \equiv \frac{l_v u}{\alpha}
\end{align}
is the P{\`e}clet number, $\alpha$ is the radiative thermal diffusivity, $N$ is the \brvs\ frequency, $a\approx 1$ and $C \approx 0.08$.
Putting this together we may write equation~\eqref{eq:nuv0} as
\begin{align}
\nu_v \approx \frac{\alpha C}{1 + \frac{u \alpha}{l_v^3 N^2}}\left(\frac{u}{N l_v}\right)^2.
\label{eq:nuv}
\end{align}
Note that this doubly-diffusive instability likely does not emerge in GCMs because it typically operates on scales much smaller than the GCM grid scale. As we shall see this instability has significant consequences for the flow speeds we obtain, particularly near the top of the atmosphere, and may explain some of the differences we see between our predictions and those of GCMs.

Inserting equations~\eqref{eq:nuh} and~\eqref{eq:nuv} into equation~\eqref{eq:powerbalanc2} we obtain
\begin{align}
	u^2\left(\frac{1}{l_h} + \frac{C}{1 + \frac{u \alpha}{l_v^3 N^2}}\left(\frac{1}{l_v}\right)\left(\frac{u \alpha}{l_v^3 N^2}\right)\right) \approx \max\left(|\boldsymbol{\Lambda}|,\left|\frac{P}{\rho}\frac{\Delta T}{T_{\rm avg}\pi R}\right|\right).
\end{align}
This may be further simplified by noting that there is no turbulent forcing in the radiative zone, so
\begin{align}
	\frac{u^2}{l_v}\left(\frac{l_v}{l_h} + \frac{C}{1 + \frac{l_v^3 N^2}{u \alpha}}\right) \approx \left|\frac{P}{\rho}\frac{\Delta T}{T_{\rm avg}\pi R}\right|.
    \label{eq:radbalance}
\end{align}

\subsubsection{Convective Zones}
\label{sec:convvisc}

In convection zones two effects may set the scale of the viscosity.
First, convection generates an eddy viscosity
\begin{align}
	\nu_h \approx \nu_l \approx \frac{1}{3} \varv_c h,
	\label{eq:convvisc}
\end{align}
where $\varv_c$ is the convection speed and
\begin{align}
	h \equiv -\frac{dr}{d\ln p}.
\end{align}
is the pressure scale height.

Additionally, and analogously to the turbulent viscosity in Section~\ref{sec:radvisc}, there is a contribution owing to the shear itself.
That is, the shear generates turbulent eddies with vertical length-scale $l_v$ and horizontal length-scale $l_h$ and velocity scale $u$.
Hence there is an additional contribution of the form
\begin{align}
	\nu_i \approx \frac{1}{3} l_i u.
\end{align}
for $i \in \{v, h\}$.
Combining this with equation~\eqref{eq:convvisc} we find that
\begin{align}
	\nu_i \approx \frac{1}{3}\left(\varv_c h + u l_i\right).
    \label{eq:nuconv}
\end{align}

A further term we must consider is the turbulent forcing $\boldsymbol{\Lambda}$.
This vanished in the case of radiative zones because the turbulence is driven by the flow, but in convection zones there is turbulence even when $u=0$.
In studies of this forcing it is usually divided into a component along the azimuthal direction, known as the $\Lambda$-effect, and one along the latitudinal direction.
The former is of order
\begin{align}
	\Lambda_\phi \approx \frac{\varv_c^2}{h}\min\left(1, \frac{h \Omega}{\varv_c}\right)
\end{align}
(\citealt{2014A&A...572L...7R}; for the rapid-rotation limit see~\citealt{doi:10.1093/mnras/sty255}),
while the latter is found to scale by various approaches as~\citep{1993A&A...276...96K,Gough2012,doi:10.1093/mnras/sty255}
\begin{align}
	\Lambda_\theta \approx \frac{\varv_c^2}{h}\min\left(1, \left(\frac{h \Omega}{\varv_c}\right)^2\right).
\end{align}
Because the azimuthal forcing is always at least as large as the meridional forcing we keep only the former.
Hence
\begin{align}
	\Lambda \approx \Lambda_\phi \approx \frac{\varv_c^2}{h}\min\left(1, \frac{h \Omega}{\varv_c}\right).
    \label{eq:forcing}
\end{align}
As we shall see the thermal forcing is usually much stronger than this, but we include this effect because the two are comparable in the solar system gas giants.

Inserting equations~\eqref{eq:nuconv} and~\eqref{eq:forcing} into equation~\eqref{eq:powerbalanc2} we obtain
\begin{align}
	\frac{u}{3}\left(\frac{\varv_c h}{l_v^2} +\frac{\varv_c h}{l_h^2} + \frac{u}{l_v} + \frac{u}{l_h}\right) \approx \max\left(\frac{\varv_c^2}{h}\min\left(1, \frac{h \Omega}{\varv_c}\right),\left|\frac{P}{\rho}\frac{\Delta T}{T_{\rm avg}\pi R}\right|\right).
    \label{eq:convbalance}
\end{align}

\subsection{Length-Scales}
\label{sec:scales}

We now seek to determine $l_v$ and $l_h$.
We expect that $l_v \approx h$ because this is the scale over which the thermodynamic properties which drive the flow change.
This is also dynamically motivated: we do not expect inertia to carry motions across regions of substantially different pressure and density.
Hence we take $l_v \approx h$.

When the planet is slowly-rotating there is only one length-scale involved in horizontal motion, namely $r$.
In this limit therefore we write $l_h \approx r$.
When the planet is rotating more quickly the inverse cascade causes energy to accumulate at large length-scales~\citep{1975JFM....69..417R}.
This has been extensively studied~\citep{Sukoriansky2007,PhysRevE.81.016315}, with the conclusion that the relevant horizontal length-scale is~\citep{2015JAtS...72.3891C}
\begin{align}
	l_h \approx l_{\rm Rhines} = 2\pi \sqrt{\frac{\varv_{\rm turb}}{|\nabla_h f|}},
    \label{eq:lRhines}
\end{align}
where $\varv_{\rm turb}$ is the turbulent velocity which we shall calculate later, $f = 2\Omega \cos\theta$ is the Coriolis parameter and $\nabla_h$ denotes the gradient in the plane of the flow.
This gradient may be evaluated locally but for our purposes its typical value suffices.
Neglecting variation in $\Omega$ we average over latitudes and obtain
\begin{align}
	\frac{1}{\pi}\int_0^\pi |\nabla_h f| \sin\theta d\theta = \frac{\Omega}{\pi}\int_0^\pi 2\left|\frac{\partial \cos\theta}{\partial \theta}\right| \sin\theta d\theta = \Omega.
\end{align}
Hence
\begin{align}
	l_h \approx 2\pi \sqrt{\frac{r\varv_{\rm turb}}{\Omega}},
\end{align}
in good agreement with the scale seen in GCMs~\citep{Liu2010}.
Note that the inverse cascade becomes relevant only when $l_{\rm Rhines} < r$, as the characteristic scale cannot be larger than the planet, so we write
\begin{align}
	l_h \approx \min\left(r, \sqrt{\frac{r \varv_{\rm turb}}{\Omega}}\right).
    \label{eq:rhineslength}
\end{align}

We must still determine $\varv_{\rm turb}$.
In convection zones, following the arguments of Section~\ref{sec:convvisc} we write
\begin{align}
	\varv_{\rm turb} \approx \varv_c + u.
    \label{eq:vturbvc}
\end{align}
This reflects the fact that both the mean flow and the convective flow contribute to the overall velocity.
We might have added them in quadrature because they are likely uncorrelated, but this form is more convenient and is good to the same level of approximation we have used elsewhere.

In radiative zones there are likewise two contributions to the turbulent velocity.
In the horizontal directions there is no stratification so the turbulence has velocity scale $u$.
In the vertical direction the turbulence acts on a length-scale~\citep{0004-637X-837-2-133}
\begin{align}
	l_{\rm turb} \approx \frac{u \alpha}{l_v^2 N^2},
\end{align}
so the viscosity implies a velocity scale
\begin{align}
	\varv_{\rm turb,v} \approx \frac{\nu_v}{l_{\rm turb}}.
\end{align}
Inserting equation~\eqref{eq:nuv} we find
\begin{align}
	\varv_{\rm turb,v} \approx u\frac{C}{1 + \frac{u \alpha}{l_v^3 N^2}}.
\end{align}
This is always less than $u$, which is the contribution from horizontal turbulence, so in radiative zones we write
\begin{align}
	\varv_{\rm turb} \approx u,
    \label{eq:vturbrad}
\end{align}
which may be viewed as the $\varv_c \rightarrow 0$ limit of equation~\eqref{eq:vturbvc}.

\subsection{Analytic Solutions}
\label{sec:summary}

Equations~\eqref{eq:radbalance},~\eqref{eq:convbalance},~\eqref{eq:rhineslength},~\eqref{eq:vturbvc} and~\eqref{eq:vturbrad} determine the flow speed in our model.
However their asymptotic behaviour and scaling are not immediately apparent.
It is useful therefore to produce an analytic solution for $u$ with appropriate breakpoints in where the flow switches from being dominated by one phenomenon to being dominated by another.
This solution also makes the numerical implementation of these equations simpler and more efficient, enabling us to study many more scenarios.

To begin define
\begin{align}
	u_0 \equiv \sqrt{l_v \max\left(|\boldsymbol{\Lambda}|,\left|\frac{P}{\rho}\frac{\Delta T}{T_{\rm avg}\pi R}\right|\right)}.
    \label{eq:u0}
\end{align}
This is the characteristic velocity scale associated with the power input.
We use the notation
\begin{align}
	\bar{x} \equiv \frac{x}{u_0}.
\end{align}
That is, an over-bar denotes a quantity which has been normalised by $u_0$.

We define the velocity scale of radiative diffusion by
\begin{align}
	\eta \equiv \frac{l_v^3 N^2}{\alpha}.
	\label{eq:eta}
\end{align}
We denote the rotation speed by
\begin{align}
	u_\Omega \equiv \Omega r,
\end{align}
and define the parameters
\begin{align}
	\lambda_0 \equiv \frac{l_v}{r}
\end{align}
and
\begin{align}
	\lambda \equiv \frac{l_v}{l_h} = \lambda_0\max\left(1,\sqrt{\frac{\bar{u}_\Omega}{\bar{\varv}_c + \bar{u}}}\right),
\label{eq:unsimp}
\end{align}
where we have inserted equations~\eqref{eq:rhineslength} and~\eqref{eq:vturbvc}.
Note that equation~\ref{eq:unsimp} applies even in radiative zones because there $\varv_c$ vanishes and therefore so does $\bar{\varv}_c$.

With these definitions, the convective power balance equation~\eqref{eq:convbalance} becomes
\begin{align}
	\frac{1}{3}\bar{u}\left(\bar{u}\left(1+\lambda\right) + \bar{\varv}_c\left(1+\lambda^2\right)\right) = 1.
    \label{eq:conv_non_dim_0}
\end{align}
Likewise in the radiative case we may write equation~\eqref{eq:radbalance} as
\begin{align}
	\bar{u}^2\left(\lambda + \frac{C}{1 + \frac{\bar{\eta}}{\bar{u}}}\right) = 1.
	\label{eq:rad_non_dim_0}
\end{align}

In Appendix~\ref{appen:cases} we extract the asymptotic behaviour of $\bar{u}$ in the extreme limits of these non-dimensional equations.
We also determine appropriate breakpoints for transitioning between different limits, so that the asymptotic forms may be used everywhere while ensuring continuity.
The resulting expressions for $u$ and criteria for determining the appropriate regime are provided in Table~\ref{tab:cases}.

\begin{table}
\begin{tabular}{lllll}
& \bf{Structure} & \bf{Criteria} &&  $\bar{u}$\\
\hline
C1& Convective & $(\bar{\varv}_c/3)^2 (1/3)^{-1} > 1$ &&  $(\bar{\varv}_c/3)^{-1}$\\

\hline

C2& Convective & $(\bar{\varv}_c/3)^2 (1/3)^{-1} < 1$ && $(1/3)^{-1/2}$\\

\hline

R1& Radiative & $\bar{\eta}^2 C < 1$ && $C^{-1/2}$\\

\hline

R2& Radiative & $\bar{\eta}^2 C > 1$ && $(\lambda_0 \sqrt{\bar{u}_\Omega})^{-2/3}$\\

&& $(\lambda_0 \sqrt{\bar{u}_\Omega})^{2} \lambda_0^{-3/2} > 1$&&\\
&& $(\lambda_0 \sqrt{\bar{u}_\Omega})^{3} (C/\bar{\eta})^{-3/2} > 1 $&&\\
\hline

R3& Radiative & $\bar{\eta}^2 C > 1$ && $\lambda_0^{-1/2}$\\

&& $(\lambda_0)^{3/2} (\lambda_0 \sqrt{\bar{u}_\Omega})^{-2} > 1$&&\\
&& $(\lambda_0)^{3} (C/\bar{\eta})^{-2} > 1 $&&\\
\hline

R4& Radiative & $\bar{\eta}^2 C > 1$ && $(C/\bar{\eta})^{-1/3}$\\

&& $(C/\bar{\eta})^{3/2}(\lambda_0 \sqrt{\bar{u}_\Omega})^{-3} > 1$&&\\
&& $(C/\bar{\eta})^{2} \lambda_0^{-3} > 1 $&&\\

\hline
\end{tabular}
\caption{Different solutions for $u$ are shown along with the limits in which they apply. The case R2 is similar to the `Coriolis' case discussed by~\citet{komacek2016}, and the remaining cases are analogous to their `Advection' case with different characteristic length-scales. Derivations of these cases are given in Appendix~\ref{appen:cases}.}
\label{tab:cases}
\end{table}

The different cases in Table~\ref{tab:cases} have clear physical interpretations.
For instance in regime C1 the convective velocity is faster than the wind, so the turbulent viscosity is dominated by convection.
In C2 by contrast the wind is faster and so dominates the dissipation.

In regime R1 the stratification of the atmosphere is weak relative to the wind.
In other words the Richardson number is small, such that turbulence may be generated by vertical shearing.
In this regime because $\lambda \ll 1$ the vertical shear is stronger than the horizontal one and hence dominates the dissipation.

In the remaining three radiative regimes the stratification of the atmosphere is strong relative to the wind, such that the flow is \textit{linearly} stable against vertical shear in the absence of thermal diffusion.
In R2 and R3 the doubly-diffusive instability is weak so horizontal shear is dominant.
The distinction between the two is that the former case is rapidly rotating, such that $l_h$ is reduced from $r$, while the latter is slowly rotating and has $l_h = r$.
Finally, in regime R4 the flow exhibits the doubly-diffusive vertical shearing instability dominates the dissipation.

Some of these cases map straightforwardly onto cases discussed by~\citet{komacek2016}.
In particular our case R3 corresponds directly to their ``Advection'' regime, and produces the same answer up to a factor of $\sqrt{\pi}$ corresponding to a different choice of length-scale.
Our cases R1 and R4 are also related to their ``Advection'' regime, but with different scalings to account for the fact that the flow is stably stratified in the vertical direction.
Case C2 is related to the same regime, but with $u^2/h$ appearing rather than $\mathcal{U}\mathcal{W}/h$ because we have computed this from a turbulent stress, which scales like the shear times the velocity, while they have used the stress of the mean flow.

\subsection{Wind Heat Flux}
\label{sec:flux}

Having determined the wind speed using Table~\ref{tab:cases} we must next determine the heat flux associated with the wind.
We assume that the kinetic energy of the wind at depth $z$ is dissipated into heat locally at that depth.

Assuming that the wind leaving from the day side arrives at the night side an amount $\Delta T$ hotter than the night side, and likewise that that leaving the night side arrives at the day side an amount $\Delta T$ cooler than the day side, we find the total heat flow per unit depth to be
\begin{align}
	\dot{Q} = 2\pi R \rho u c_{\rm p} \Delta T,
\end{align}
where the factor of $2\pi R \rho u$ is the unsigned mass flux per unit depth~\footnote{The signed mass flux is zero because we have assumed the system to be in steady state, but because the temperature of the material is correlated with its direction of travel the unsigned flux is the one which matters.} and $c_{\rm p}$ is the specific heat capacity at constant pressure.
The heating per unit mass on the night side is therefore
\begin{align}
	\epsilon = \frac{\dot{Q}}{4\pi R^2 \rho} = c_{\rm p} \Delta T \left(\frac{u}{2 R}\right),
    \label{eq:epsilon0}
\end{align}
and the cooling per unit mass on the day side is the same.

As a further simplification it is useful to note that for an ideal gas
\begin{align}
	c_{\rm p} T = \frac{v_s^2}{\gamma-1},
\end{align}
where $v_s$ is the adiabatic speed of sound.
With this, equation~\eqref{eq:epsilon0} may be written as
\begin{align}
	\epsilon = \frac{\dot{Q}}{4\pi R^2 \rho} = v_s^2\left(\frac{\Delta T}{T}\right) \left(\frac{u}{2(\gamma-1) R}\right).
\end{align}

\subsection{Efficiency Factor}
\label{sec:efficiency}

The assumption underlying equation~\eqref{eq:epsilon0} is that the material which leaves the day side reaches the night side at $T_{\rm day}$, and likewise that material leaving the night side reaches the day side at $T_{\rm night}$.
This is not true, and so we must correct for radiative losses en route.

As material travels from the day side to the night side, heat is radiated to the surroundings and out of the atmosphere into space.
The heat radiated to the surroundings is carried by the flow still and so is not lost to the wind flux.
On the other hand the heat radiated into space is lost.
The heat lost in this manner before the wind crosses from the day side to the night side is accounted for in the day side flux, and that lost on the night side is accounted for in the night side flux.
Hence in the context of our model it suffices to simply reduce the wind flux from one side to the other to account for these losses.

Because losses are only incurred at low optical depth, the loss factor is expected to scale as $e^{-\tau}$, where $\tau$ is the optical depth.
Furthermore because losses are incurred by radiation acting as the wind circles the planet we expect the losses to be proportional to the ratio between the wind-crossing time-scale and radiative thermal time-scale.
This ratio is
\begin{align}
\frac{t_{\rm wind}}{t_{\rm rad}} = \frac{\pi R}{u t_{\rm rad}},
\label{eq:ratio}
\end{align}
where
\begin{align}
	 t_{\rm rad} = \frac{h \rho c_p T}{F},
\label{eq:trad}
\end{align}
is the local thermal time-scale evaluated over one pressure scale-height and $F$ is the frequency-summed heat flux.
When the ratio in equation~\eqref{eq:ratio} is small the radiative losses are proportionately small.
When this ratio is large the radiative losses are of order the entire flux.
Hence we write the efficiency factor as
\begin{align}
	f = 1 - e^{-\tau} \min\left(1,\frac{t_{\rm wind}}{t_{\rm rad}}\right).
\end{align}
Here, $\tau$ refers to the optical depth derived from the Rosseland mean opacity. With this the flux transported between the two sides of the atmosphere is
\begin{align}
F_\mathrm{wind} = f \epsilon.
\end{align}
We define the quantity $D$ as the flux per unit radius,
\begin{align}
D = f \rho \epsilon = f \rho v_s^2 \frac{\Delta T}{T} \frac{u}{2(\gamma-1)R}.
\label{eq:D_calculation}
\end{align}
This quantity used to modify the radiative-convective equilibrium equations for the day and night sides (see Appendix~\ref{sec:rt_rceqm}).

\section{Winds in Radiative Transfer and Radiative-Convective Equilibrium}\label{sec: GENESIS modifying}

We incorporate the wind heat flux into the one-dimensional radiative-convective (thermal) equilibrium model in the GENESIS code~\citep{gandhi2017} by adapting the prescription of \citet{burrows2008}. The modified GENESIS equations are discussed in Appendix~\ref{sec:rt_rceqm}. For the radiative transfer equation, the boundary condition at the top of the atmosphere is different between the two sides because the night side receives no incident flux (see Figure~\ref{fig:sketch}). Otherwise the radiative transfer equation is identical for both sides of the atmosphere.

Radiative-convective equilibrium ensures that the energy flowing into a given region of the atmosphere is matched by the energy that exits that region. The radiative-convective equilibrium equations in GENESIS are determined by the incident flux from the star and the emergent flux from the planet's internal heat. Our change now is to include an additional depth dependent wind flux, $D$, which removes flux from the day side and adds an equivalent flux to the night side. The details of the procedure are given in Appendix~\ref{sec:rt_rceqm}.

\subsection{Procedural Overview}\label{sec:overview}

We now summarise the steps taken to compute the day and night side atmosphere using the wind model. We begin with an initial solution for the day and night side temperature profiles. We use the converged equilibrium solution with half of the stellar flux incident on the day side and half on the night side, meaning that the two sides initially have identical profiles. This initial starting condition therefore starts with no wind as $F_\mathrm{wind} = 0$ when $\Delta T =0$. In the following steps the day side receives the full incident flux from the star and night side receives none. This drives a temperature difference between the day and night sides and thus generates a wind flux. We have verified that our model does converge to the same solution regardless of the initial starting profiles, but using this initial solution ensures that fewer iterations are required as this is close to the converged solution for the deep atmosphere. We have also verified the convergence of our wind model by varying both the internal flux temperature between 100-300~K and the bottom pressure that we model to.

Once we have an initial solution, we then:
\begin{enumerate}
\item Compute the wind velocity in each layer of the atmosphere given the temperature on the day and the night side from Table~\ref{tab:cases}.
\item Calculate a flux $F_\mathrm{wind}$ associated with the wind in each layer of the atmosphere. This determines the energy transported from the day side to the night side.
\item Calculate the absorption and scattering coefficients of each side of the planet.
\item Compute the wind heating term $D$ from equation~\ref{eq:D_calculation}.
\item Use a modified Newton-Raphson solver (i.e. the complete linearisation method) to determine the correction to the temperature on the day side from the radiative transfer and radiative-convective equilibrium equations (see Appendix~\ref{sec:rt_rceqm}).
\item Repeat the above step for the night side.
\item Update the temperature profile in each layer of the atmosphere for the day and the night side. These new profiles are used in the next iteration to determine the wind flux in step (ii).
\item Repeat the steps (i)-(vii) above until the relative change in the temperature profile is less than a fixed cutoff for both sides of the planet.
\end{enumerate}

We assume for this study that the day and night sides are each in chemical equilibrium \citep{heng2016, gandhi2017}. This  means that we have assumed that the timescale for chemical reactions $\tau_\mathrm{chem}$ is much less than the advective timescale $\tau_\mathrm{adv}$ for the wind to transport material between the two sides of the atmosphere. We leave the exploration of chemical disequilibrium on hot Jupiters \citep[see e.g.][]{cooper2006} for future work.

For planets with temperatures in excess of 2000~K we also include the effect of thermal dissociation of H$_2$O, TiO, H$_2$ and H- \citep{parmentier2018, gandhi2020_h-} in the atmosphere and the recombination/dissociation energy from the dissociation of H$_2$ \citep{komacek2018} into the wind flux. This is done by incorporating an additional $Q_\mathrm{recomb}$ term on the right hand side of equation~\ref{eq:epsilon0}.

We use the most complete available high temperature line lists for the computation of the opacity and spectra. The line lists for H$_2$O \citep{polyansky2018}, HCN \citep{barber2014}, NH$_3$ \citep{coles2019} and C$_2$H$_2$ \citep{chubb2020} are sourced from the ExoMol database \citep{tennyson2016} and that for CO, CO$_2$ and CH$_4$ and from the HITEMP database \citep{rothman2010, li2015, hargreaves2020}. We broaden each line on a grid of temperatures and pressures spanning typical photospheric conditions for such planets \citep{gandhi2017}, using H$_2$/He broadening coefficients where available \citep[see][for further details]{gandhi2020_cs}. We also introduce collisionally induced absorption from the HITRAN database for the H$_2$/He rich atmospheres of these hot Jupiters \citep{richard2012}. For models of ultra-hot Jupiters with temperatures in excess of 2000~K we also introduce opacity from TiO \citep{mckemmish2019}, H- \citep{bell1987, john1988} and Fe \citep{kramida2018}. 

We assume the planets are tidally locked and therefore that the rotation rate is equal to its period. We model atmosphere in hydrostatic and local thermodynamic equilibrium for both the day and night side, under the assumption of an ideal gas. The atmosphere to discretised into 100 layers evenly spaced in $\log(P)$ between $10^3-10^{-5}$~bar, with 10,000 evenly spaced frequency points between 0.4-30~$\mu$m for both the day side and night side. Further details of the model setup can be found in \citet{gandhi2017}. 

The stellar flux incident on the day side is
\begin{align}
H_\mathrm{ext} = \frac{F_\mathrm{star}}{4\pi}\frac{R_\mathrm{star}^2}{a^2}.
\end{align}
Here, $a$ is the semi-major axis of the orbit and the stellar radius is $R_\mathrm{star}$. $F_\mathrm{star}$ is calculated from the Kurucz model spectra \citep{Kurucz_1979_paper, kurucz_model} and varies with the temperature, metallicity and $\log(g)$ of the star. Assuming $F_\mathrm{star}$ can be written $\sigma_R T_\mathrm{eff,star}^4$, where $\sigma_R$ is the Stefan-Boltzmann constant, the equilibrium temperature $T_\mathrm{eq} = T_\mathrm{eff,star} \sqrt{R_\mathrm{star}/2a}$ for a planet with full redistribution.

\section{Validation}\label{sec:validation}

\begin{figure*}
  \centering
    \includegraphics[width=\textwidth,trim=0 0cm 0 0,clip]{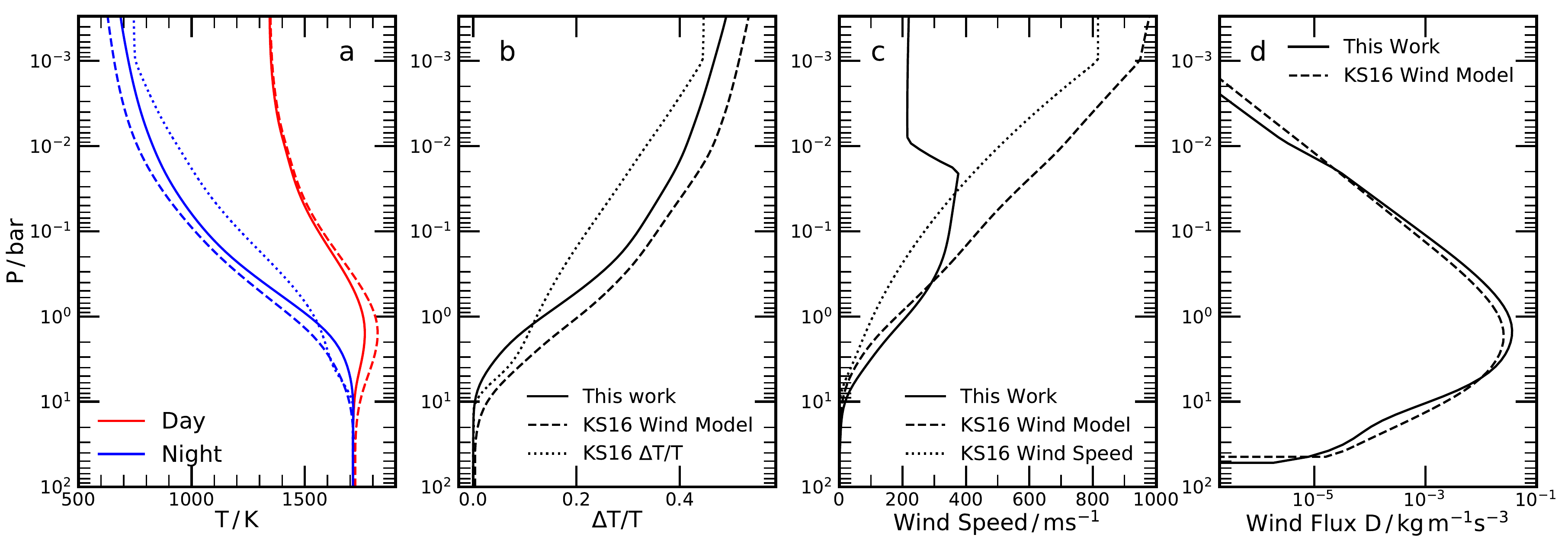}
  \caption{Comparison of the temperature profiles, wind speeds and wind fluxes from the wind models in our work and \citet{komacek2016}. Panel a shows the temperature and the panel b the ratio $\mathrm{\Delta T/T}$ as a function of pressure for the planet WASP-43~b. Panel c shows the wind speeds and panel d the wind flux $D$ calculated from equation~\ref{eq:D_calculation}. We also show the temperature and wind profiles calculated from the full analytic prescription in \citet{komacek2016} for comparison with the dotted lines. Note that the wind flux $D$ is not used in their analytic prescription.}
  \label{fig:wind_compare}
\end{figure*}

We now validate both our wind model and its use in GENESIS against analytic results by~\citet{komacek2016} (see also \citet{zhang2017, komacek2017}).
We do this with three separate calculations for the exoplanet WASP-43~b using the system parameters given in Table~\ref{tab:system_params}.
The results are shown in Figure~\ref{fig:wind_compare}.
The first is a calculation done with our wind model and radiative transfer and thermal equilibrium handled by GENESIS, as discussed in Section~\ref{sec:overview}.
The second is a calculation done with the analytic wind model of~\citet{komacek2016}, hereinafter the KS16 wind model, combined with radiative transfer and thermal equilibrium handled by GENESIS.
Finally, we include the full results of the KS16 model, hereinafter KS16 $\Delta T/T$, in which both the wind and radiative transfer are handled analytically using their prescription in panels b and c of Figure~\ref{fig:wind_compare}.

The three calculations agree at the $20\%$ level on the day-night temperature difference, both in magnitude and in the dependence on depth.
The first and second calculations agree particularly well, while the third shows deviations near the top of the atmosphere due to greater differences in the assumptions on the day and night sides.
In particular, for the first and second calculations we determine the thermal time scale using equation~\ref{eq:trad}, which is more physically motivated than the analytic scaling KS16 use.

In the deep atmosphere (P$\gtrsim10$~bar) we see that the two sides begin to converge towards the same temperature. This is because the wind transport is very efficient at such pressures, and thus very little of the incident stellar flux that is deposited into the day side is able to escape in order to significantly cool the night side. This isothermal region of the atmosphere does still have a small temperature difference between the two sides, and it is only in the convective regions of the very deep atmosphere that the two sides will in fact be equal. This is also in agreement with GCMs \citep[e.g.][]{kataria2015} and the KS16 model.

On the other hand, \citet{burrows2008} has shown that such an isothermal zone may have a much larger temperature difference between the two sides. 
This is because they take the heat redistribution to be limited to a fixed region of the atmosphere, whereas in our model the heat redistribution actually becomes more efficient per unit $\Delta T/T$ with increasing depth due to the increasing specific thermal energy (Eq.~\ref{eq:D_calculation} and the right-most panel of Figure~\ref{fig:wind_compare}).

We see larger differences in wind speeds between the three calculations.
Deeper than $0.1\,\mathrm{bar}$ all three agree to better than $30\%$.
In shallower regions the second and third calculations, both of which use the KS16 wind model, continue to agree well while our wind model predicts speeds that range from $2$ to $5$ times slower than the others.
There are two reasons for this.

Firstly, this planet is mostly in case R2 in our model and the ``Coriolis'' regime in the KS16 model.
While these regimes are conceptually similar, they exhibit different scalings because we and KS16 treat rotation very differently.
Because our model is based on balancing the work done by the temperature gradient with the energy dissipated by turbulence, and because the Coriolis force does no work, this force never appears directly in our equations\footnote{We explicitly include the Coriolis term in the beginning and then drop it in going from equation~\eqref{eq:yes_coriolis} to equation~\eqref{eq:no_coriolis}}.
Instead, a dependence on the rotation rate enters through the horizontal length-scale $l_h$, which is set by the Rhines scaling law (equation~\eqref{eq:lRhines}).
By contrast in the KS16 model the Coriolis force is used directly to balance the pressure gradient which drives the flow.
Given that the flow undergoes significant dissipation even while circling the planet once (i.e. the dissipation time-scale is of order or less than $R/u$) we favour our implementation of rotational effects, though future numerical simulations should be able to provide stronger evidence one way or the other.

Secondly, we have assumed that the vertical length scale of the flow is on the order of one pressure scale height, whereas the flow is seen to be coherent over a greater scale in GCMs \citep[e.g.][]{kataria2015}. This results in us overestimating the turbulent dissipation and therefore underestimating the flow speed, particularly in the upper atmosphere where we see the largest discrepancy.
By modifying the coherence length we can make our results match those of GCMs even more closely, but we leave a precise calibration of this to the future.

In Section~\ref{sec:wasp43 winds} we further compare our results with those of GCMs and typically find good agreement, though only for specific values of the GCM drag time-scale, suggesting that our wind model can be interpreted as giving a scheme to compute $\tau_{\rm drag}$.
We shall discuss this point further in that section.

Figure~\ref{fig:wind_compare} also shows the wind flux $D$ calculated from equation~\ref{eq:D_calculation} for our wind model and the KS16 wind model. These agree well for all pressures and show that the strongest flux occurs at P$\sim$1~bar, where the majority of the stellar flux is deposited on the day side. At higher pressures $D$ decreases as the temperature difference between the day and night sides and the wind speeds decrease. At pressures $\lesssim$1~bar, a significant portion of the deposited stellar flux is re-radiated out to space given the low optical depth and low pressure. Hence the wind does not transfer a significant flux to the night side at such pressures. At higher pressures (P$\gtrsim10$~bar), the flux also drops because the day and night sides have a much smaller temperature difference.

\section{Results and Discussion}\label{sec:results}

In this section we explore models of various hot Jupiters over a wide range of temperatures. We begin by modelling two cases, WASP-76~b, an ultra-hot Jupiter which showed a thermal inversion and Fe condensation on the night side \citep{ehrenreich2020, fu2020} and WASP-43~b, which has high precision thermal phase curves \citep{stevenson2014, stevenson2017} as well as previous GCM analyses \citep{kataria2015} to compare. We additionally use our HyDRA retrieval framework to constrain deviations in the temperature profile from the radiative-convective equilibrium wind model for WASP-43~b. We finally explore a wide grid of hot Jupiters with equilibrium temperatures ranging between 1000-3000~K to determine how the temperature difference between the day and night is affected, and compare this grid of models to theoretical predictions and measurements from real systems \citep{keating2019}.

\subsection{WASP-76~b}\label{sec:wasp76}

\begin{figure*}
  \centering
    \includegraphics[width=\textwidth,trim=0cm 0cm 0cm 0,clip]{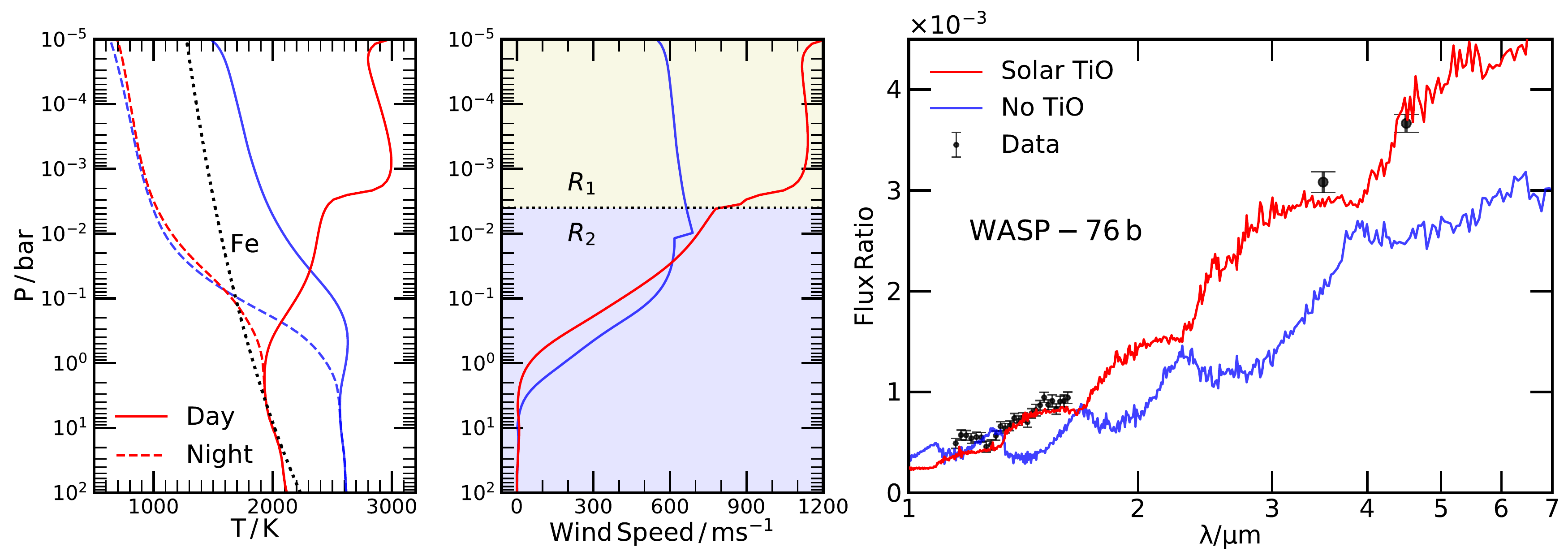}
  \caption{Effect of TiO on the atmospheric profile and day side emission spectrum of WASP-76~b. The left panel shows the temperature profile, with the solid and dashed lines indicating the day and night side profiles respectively. We also include the Fe condensation curve from \citet{visscher2010} to show the day side photosphere is hotter and the night side photosphere is cooler than the curve for both of the cases with and without TiO. The middle panel shows the wind speed as a function of pressure and is labelled with the relevant regimes from Table~\ref{tab:cases}. The right panel shows the corresponding day side thermal emission spectrum for each model. Also shown are the HST WFC3 and Spitzer photometric thermal emission observations of WASP-76~b \citep{fu2020}.}
  \label{fig:wasp76}
\end{figure*}

\begin{table}
\caption{Parameters for the WASP-76 and WASP-43 systems used to generate our models. These were chosen from \citet{fu2020} and \citet{kataria2015} respectively for consistency. We also assume an internal heat flux temperature T$_\mathrm{int} = 100$~K for both planets.}
\begin{tabular}{ll|l|l}
\multicolumn{2}{l}{\bf System Parameters}& \bf{WASP-76} & \bf{WASP-43}\\
\hline
\bf{Star} & R$_\mathrm{star}$/ R$_\odot$ & 1.74& 0.667\\
& $\log(g_{\mathrm{star}}/\,\mathrm{cm s^{-2}})$ & 4.12 & 4.65 \\
& T$_{\mathrm{eff,star}}$/ K & 6360 & 4400\\
& Z$_{\mathrm{star}}$ & 0.20 & -0.05\\
\hline
\bf{Planet} & R$_\mathrm{planet}$/ R$_\mathrm{J}$ & 1.84 & 1.036\\
& M$_\mathrm{planet}$/ M$_\mathrm{J}$ & 0.91 & 2.03\\
& a/ AU & 0.0330 & 0.01526\\
& T$_\mathrm{int}$/ K & 100 & 100 \\
\end{tabular}
\label{tab:system_params}
\end{table}

The ultra-hot Jupiter WASP-76~b has an equilibrium temperature of $\sim$2200~K~\citep{west2016}, and low resolution HST WFC3 and Spitzer observations reveal that it has a stratosphere \citep{fu2020}. To model this planet we explore two cases, one with and one without gaseous TiO, a molecule with strong optical opacity known to cause thermal inversions.
We also include the thermal dissociation of TiO and H$_2$O as well as opacity from H-, both of which have been demonstrated to be important for such hot exoplanets \citep[e.g.][]{arcangeli2018, parmentier2018}.
We further explore Fe condensation on the night side of WASP-76~b which has been observed from recent high resolution observations of the terminator with ESPRESSO/VLT \citep{ehrenreich2020}. Our assumed stellar and planetary parameters are shown in Table~\ref{tab:system_params}.

The pressure-temperature (P-T) profiles for the day and night side are shown in the left panel of Figure~\ref{fig:wasp76}. For both cases with and without TiO, the deepest layers of the atmosphere, P$\gtrsim$3~bar, show no temperature differences between the two sides of the atmosphere. At these high pressures, the efficiency factor of the wind transport is $\sim1$ and the optical depth $\tau>>1$. Thus the day and night sides are equal in temperature given that very little flux is lost out to space as the wind transports it to the night side. The temperature difference between the two sides of the atmosphere begins to increase at pressures below this because the wind loses more heat to space at low optical depth, resulting in less net heat transfer.

\subsubsection{Thermal Inversions}

Stratospheres, or thermal inversions, significantly alter the emission spectrum of the atmosphere by producing emission features in the infrared from prominent spectrally active species such as H$_2$O. A number of observations of hot Jupiters have shown evidence for these features, suggesting that they are relatively common~\citep[e.g.][]{haynes2015, sheppard2017, arcangeli2018, mikal-evans2020}.

Inversions on the atmospheres of hot Jupiters have been explained as owing to species such as TiO \citep{hubeny2003, fortney2008, spiegel2009, piette2020}, which possess strong cross sections at visible wavelengths. Even trace amounts of these species can lead to significant changes in the emission spectra, so understanding their abundance and effect in the atmosphere is paramount.

In our model the presence of TiO at solar abundance \citep{asplund2009} results in a thermal inversion and thus emission features in the dayside spectrum as shown in Figure~\ref{fig:wasp76}, which offers a better fit to the observations than a profile without TiO which is accordingly lacks an inversion. The inverted temperature profile is also consistent with the retrievals performed by \citet{fu2020}.

At pressures $\lesssim3\times10^{-3}$~bar the inversion becomes much stronger as the thermal dissociation of H$_2$O prevents the upper layers of the atmosphere from cooling effectively. The strong absorption of the stellar flux in the upper atmosphere also results in pressures $\gtrsim3\times10^{-2}$~bar being shielded and thus cooler than the case without TiO.

Many other species are also capable of producing thermal inversions in hot Jupiters \citep[e.g.][]{molliere2015, gandhi2019}. As such, multiple of these refractory species may be present and add to the thermal inversion. In our models of WASP-76~b we include opacity from gaseous Fe, a species predicted to cause inversions on ultra-hot Jupiters \citep{lothringer2018}, but for the temperatures that we are considering the inversion is dominated by the presence of TiO. In addition, we see a small inversion at P$\sim1$~bar for the non-inverted case, but this is below the infrared photosphere and thus not observable in the spectrum.

\subsubsection{Condensation of Fe}
Figure~\ref{fig:wasp76} shows the P-T profile for the two equilibrium models with and without TiO along with the condensation curve of Fe from \citet{visscher2010}.
We see that both of the models have a day side that is hot enough for Fe to be gaseous. In addition, both models also show a night side that is cool enough for Fe to condense for $P\lesssim0.1$~bar, consistent with the rainout of Fe seen by \citet{ehrenreich2020} in high resolution observations. 

While both the models with and without TiO produce Fe condensation on the night side, the one with TiO is able to more closely match the low resolution observations. For this model, the temperature profile of the deep atmosphere (P$\gtrsim1$~bar) lies close to the condensation curve of Fe, so Fe condensation may also occur at these high pressures depending on the details of heating and cooling in the deep atmosphere.

\subsubsection{Wind Speeds}

The strongest winds in the upper atmosphere ($\mathrm{P}\lesssim2\times10^{-2}$~bar) appear in the model with TiO because the day-night temperature difference is larger in this case. At the very top of the atmosphere our wind model predicts a speed $\sim$1.2~km/s. The model without TiO on the other hand has a higher wind speed at pressures $\gtrsim2\times10^{-2}$~bar. This is because the lack of TiO reduces the optical depth in the visible, causing the stellar flux to be absorbed at higher pressures and driving the wind more strongly there. As a result, the wind speed approaches 0 above $\sim1$~bar for the case with TiO, but the case without TiO only approaches 0 above P$\sim4$~bar. This may also be seen in the temperature difference between the day and night sides, which decreases more quickly at higher pressures with TiO than without it.

\subsection{WASP-43~b}\label{sec:wasp43}

\begin{figure*}
  \centering
    \includegraphics[width=\textwidth,trim=0 0cm 0 0,clip]{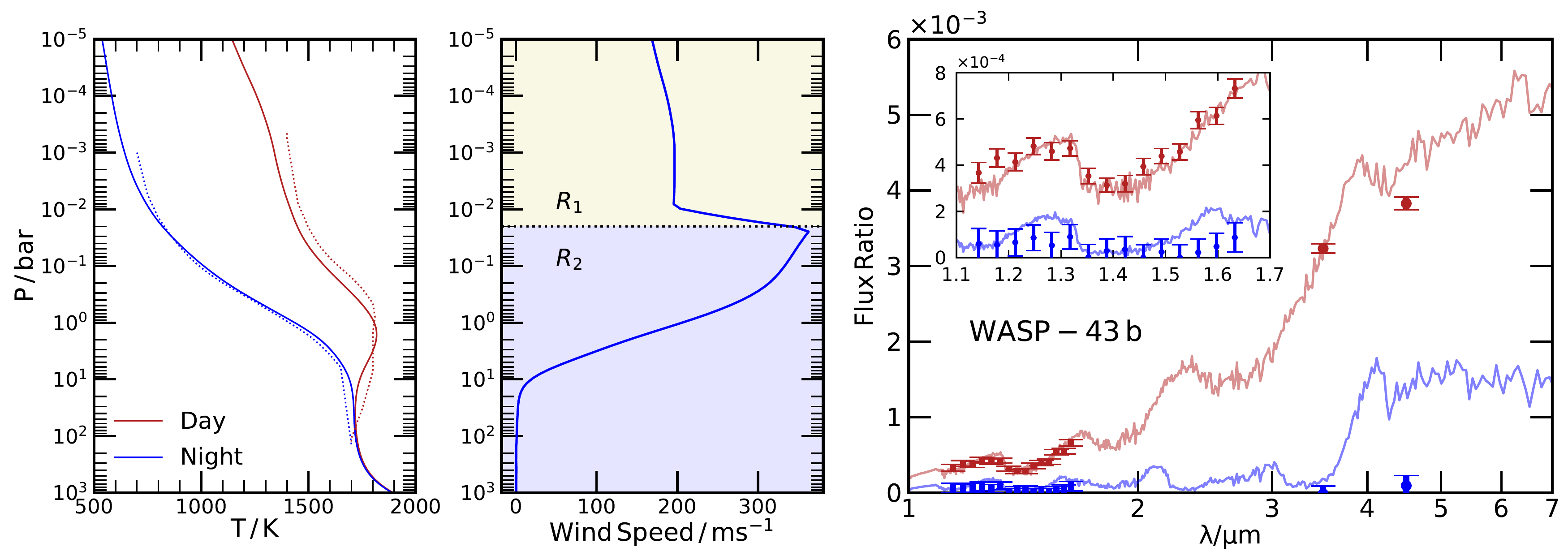}
  \caption{Atmospheric properties for WASP-43~b using the GENESIS forward model with winds. The left panel shows the radiative-convective equilibrium day and night side pressure-temperature profile for our model with the solid lines. The dashed lines show the temperature profiles for the day side (0$^\circ$ longitude) and night side (180$^\circ$ longitude) from \citet{kataria2015}, weighted by the cosine of the latitude. The middle panel shows the wind speed as a function of pressure and is labelled with the relevant regimes from Table~\ref{tab:cases}. The right panel shows the corresponding emergent spectrum for both sides of the planet, as well as day and night side observations for WASP-43~b using thermal phase curves \citep{stevenson2014, stevenson2017}, with the inset showing the HST WFC3 spectral range.}
  \label{fig:wasp43}
\end{figure*}

\begin{figure*}
  \centering
    \includegraphics[width=\textwidth,trim=0 0cm 0 0,clip]{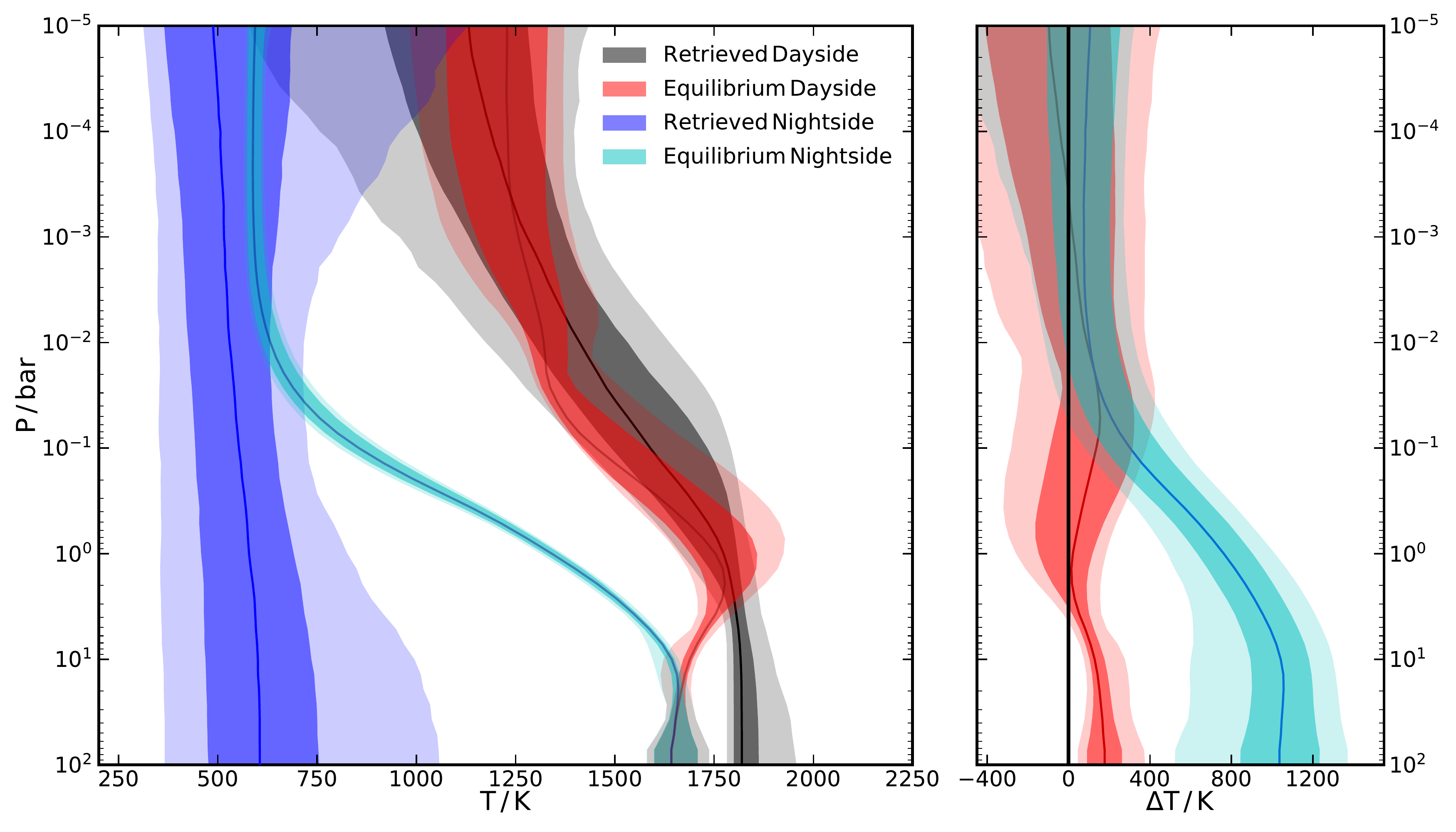}
  \caption{Retrieved and radiative-convective equilibrium pressure-temperature profiles of WASP-43~b, shown for the day and night side atmosphere. The retrievals on the day and night side emission spectra \citep[obtained from][]{stevenson2014} and the radiative-convective disequilibrium calculations were carried out using the HyDRA retrieval framework \citep{gandhi2018}. The day and night side equilibrium models incorporate the wind model as discussed in Section~\ref{sec:theory}, and is computed using the retrieved chemistry. The uncertainty in the equilibrium models is caused by uncertainties in the retrieved atmospheric chemistry. }
  \label{fig:wasp43_ret}
\end{figure*}

\subsubsection{Day and Night Side Emission Spectrum}

The day side and night side emission spectrum for our forward model is shown in the right panel of Figure~\ref{fig:wasp43}, along with HST WFC3 and Spitzer photometric observations by \cite{stevenson2014}. We see good agreement between the observations and our wind model for both the day and night side. The day side atmosphere clearly shows the absorption feature in the WFC3 range from the non-inverted temperature profile. The Spitzer observations also show a good fit to the radiative and chemical equilibrium profile of the wind model. 

The night side is cooler and thus has a significantly lower planet/star flux ratio to the day side. There is a weak absorption feature in the night side but the spectrum is largely featureless and the temperature is too cool for us to place significant constraints given the measurement uncertainties. We note that the Spitzer 4.5$\mu$m measurement is not consistent with our model, as the model spectrum shows significantly greater flux than the observations. This is a well known feature of night side emission spectra and may be explained through cloud formation on the night side due to the cooler temperatures \citep[e.g.][]{steinrueck2018}.

\subsubsection{Atmospheric Temperature Profile}

The atmospheric temperature profiles in radiative-convective and chemical equilibrium from our wind model, shown in Figure~\ref{fig:wasp43}, are in good agreement with GCMs by \cite{kataria2015} in chemical equilibrium at solar metallicity (see Figure~\ref{fig:wasp43}). Note that for the comparisons with \citet{kataria2015} we show the 0$^\circ$ and 180$^\circ$ longitudes for the day and night sides respectively. The offset hotspot means that these are not the coolest or hottest temperatures in the atmosphere. Those simulations of WASP-43~b predict that in the deep atmosphere the temperature on both the day and night side is $\sim$1700~K, which decreases to $\sim$1300~K at the top of the atmosphere for the day side and $\sim$700~K for the night side, in agreement with our model. Our model is also able to capture the small inversion they see at $\sim$1~bar on the day side. This is an encouraging sign that our relatively simple wind model is able to produce atmospheric profiles that are similar to those predicted from much more complex GCM calculations.

We also compare our equilibrium wind model to observations by retrieving the day and night side temperature profile from the HST and Spitzer data \citep{stevenson2014, stevenson2017}. The retrieved day and night side temperatures from HyDRA \citep{gandhi2018} are shown in Figure~\ref{fig:wasp43_ret}.
Using samples drawn from the retrieved chemical composition distribution we additionally computed a range of equilibrium pressure-temperature profiles with our wind model\footnote{Note that because we use the retrieved chemistry these models are not in chemical equilibrium, unlike the model shown in Figure~\ref{fig:wasp43}.}.

The photosphere and top of the atmosphere, P$\lesssim10^{-1}$~bar, are in good agreement between the equilibrium wind model and the retrieval for both the day and night side. Below the photosphere at high pressures the atmosphere is not observable, particularly for the night side, where previous work has indicated the presence of a cloud deck at $\lesssim0.2$~bar \citep{irwin2020}. Thus in the absence of constraints the retrieval sets the temperature to an isotherm.
By contrast our wind model predicts that the temperature varies quite significantly, particularly for the night side, as we go deeper into the atmosphere. The two sides eventually reach an equal day and night side temperature of $\sim$1700~K at P$\gtrsim$10~bar.
This shows that our wind model is able to derive a non-trivial atmospheric profile even at depths which are not directly observable by imposing physical constraints on the heat flux on the day and night sides.

\subsubsection{Wind Speeds}\label{sec:wasp43 winds}

\begin{figure*}
  \centering
    \includegraphics[width=\textwidth,trim=0cm 0cm 0cm 0,clip]{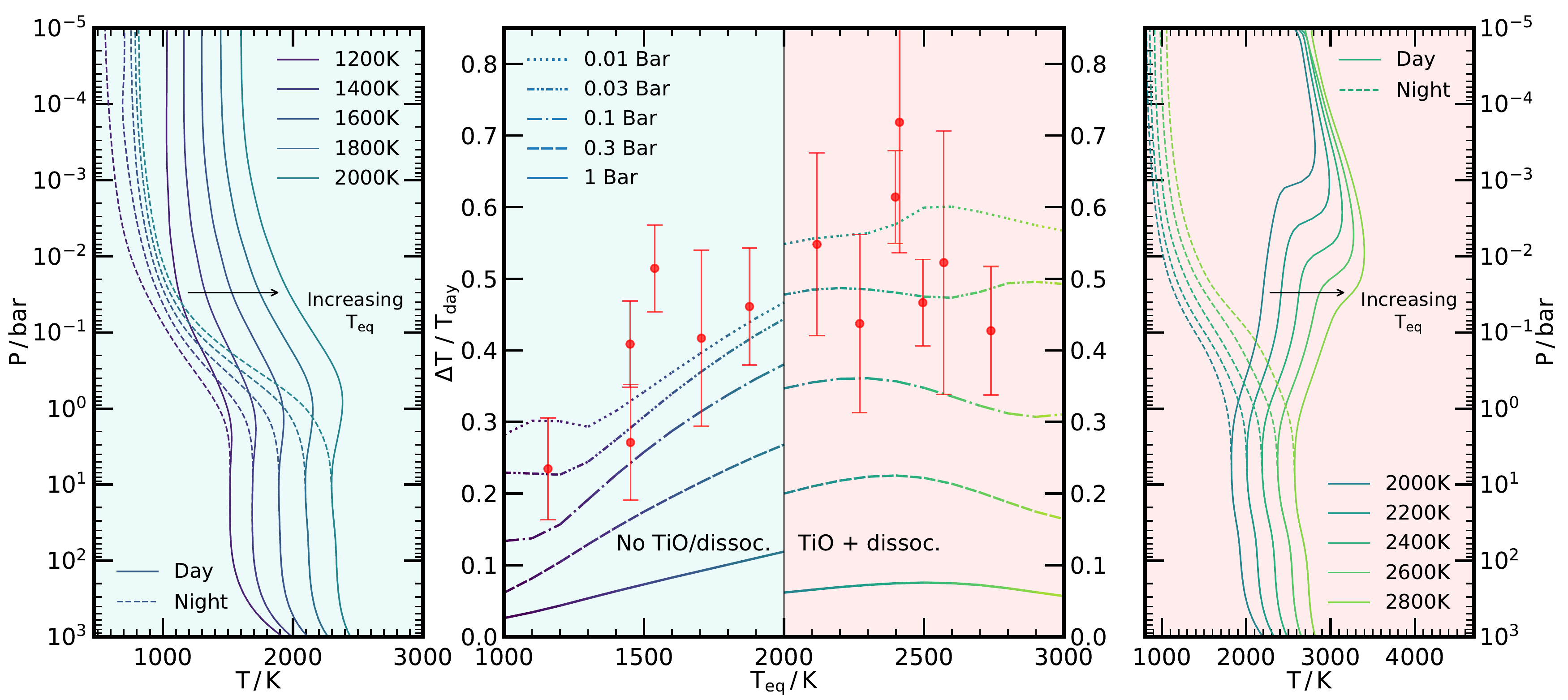}
  \caption{Variation of the temperature difference between the day and night side with equilibrium temperature, T$_\mathrm{eq}$. The middle panel shows the temperature difference relative to the day side temperature, $\Delta T/T_\mathrm{day}$, as a function of equilibrium temperature for various pressures. Equilibrium temperatures greater than 2000~K include absorption from TiO as it is expected to be gaseous, which results in an increase to the optical opacity and thus a thermal inversion. Above T$_\mathrm{eq}=2000$~K, we also include the effect of H- opacity and thermal dissociation of H$_2$O, TiO, H- and H$_2$ \citep{gandhi2020_h-}. The red markers show the observed values of $\Delta T/T_\mathrm{day}$ for real systems \citep{keating2019} along with their associated uncertainty. The left panel shows example atmospheric temperature profiles for a subset of the equilibrium models run without TiO/dissociation, with the solid and dashed lines indicating the day and night side respectively. The right panel shows a subset of the temperature profiles for the models with TiO and dissociation.}
  \label{fig:dt_t}
\end{figure*}

\begin{figure}
  \centering
    \includegraphics[width=\columnwidth,trim=0cm 0cm 0cm 0,clip]{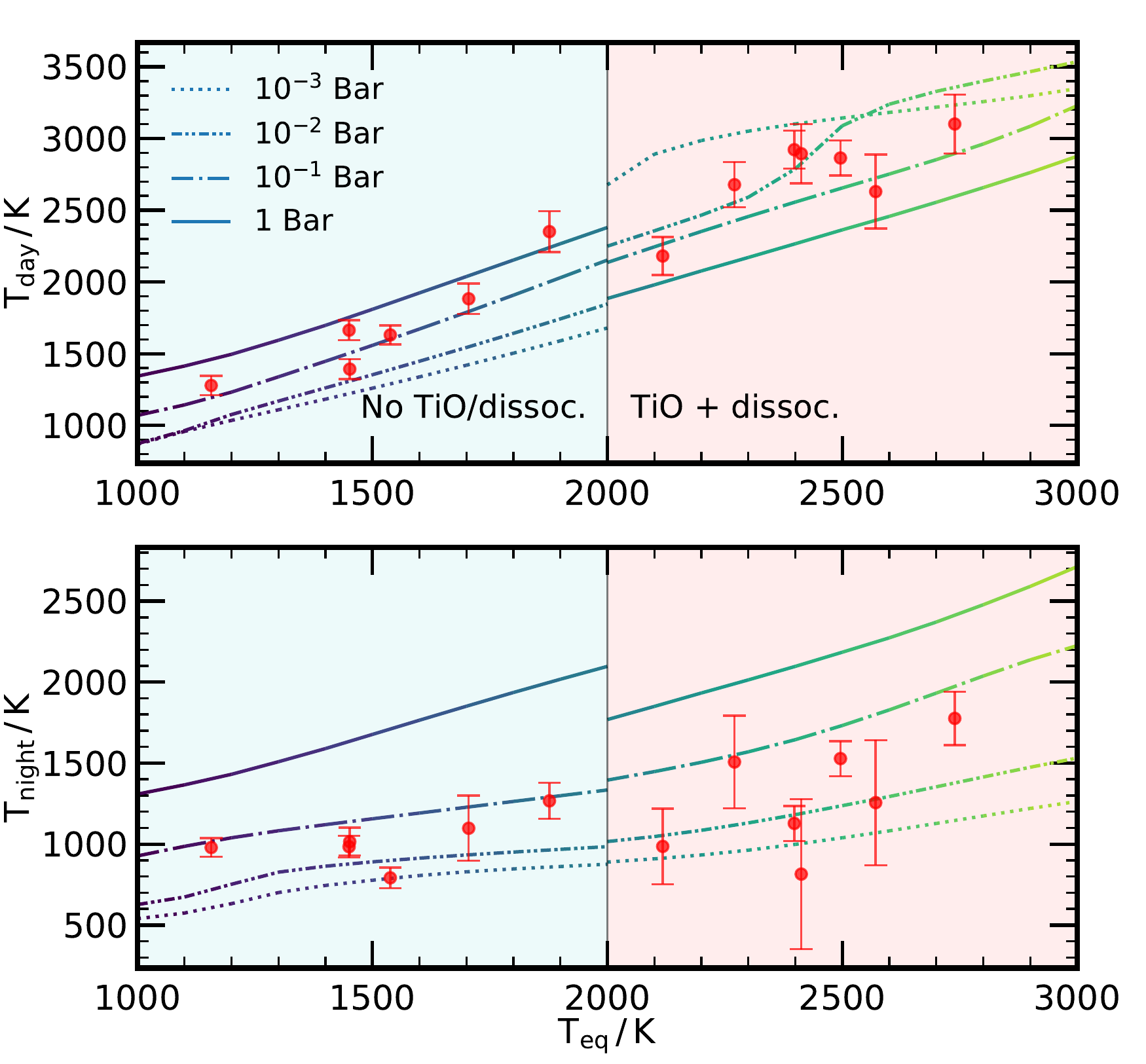}
  \caption{Day and night side temperatures as a function of equilibrium temperature for various pressures. The top panel shows the day side temperature and the bottom panel the night side temperature. We also plot the observed values of $T_\mathrm{day}$ and $T_\mathrm{night}$ for real systems \citep{keating2019} with the red markers along with their associated uncertainty. Equilibrium temperatures beyond 2000~K include opacity from TiO as well as thermal dissociation.}
  \label{fig:td_tn}
\end{figure}

The variation of the wind speed with atmospheric depth is shown in the middle panel of Figure~\ref{fig:wasp43}. The peak wind speed occurs in the photosphere, where the speed is $\sim$400m/s. This is expected given that the bulk of the incident radiation is absorbed in this region of the atmosphere, and thus will drive the strongest flux. The region in the atmosphere at P$\gtrsim2\times10^{-2}$~bar is in the radiative R2 regime as shown in Table~\ref{tab:cases}, where the flow is banded, similar to that seen for Jupiter.

At higher altitudes, the flow turns to the R1 regime, similar to the regimes seen for WASP-76~b (see Section~\ref{sec:wasp76}). The wind speeds we predict are in good agreement with those obtained with GCMs which include drag time-scales of order $10^{4}{\rm s}$ \citep{kataria2015}, though we do under-predict speeds in the upper atmosphere as noted in Section~\ref{sec:validation}. Nonetheless, we find good agreement with the predicted temperature profiles of GCMs regardless of the drag scale (see Figure~\ref{fig:wasp43}). This is in part because we predict similar heat transport for a variety of different circulation rates.

Comparison with~\citet{kataria2015} indicates that our model permits somewhat sharper changes in wind speed with depth than are seen in GCMs.
This is because our calculation of the wind speed is based only on the local properties of a single layer of the atmosphere, and so does not impose any smoothness condition beyond those imposed by e.g. thermal and hydrostatic equilibrium.
As a result the characteristic length-scale for the wind speed to change is of order the shorter of the pressure and temperature scale heights.

In GCMs, by contrast the structure of the flow persists over a longer distance, which smooths over such sharp features.
This additionally means that the effective dissipation is less in GCMs than we have assumed, which should result in greater velocities.
This is in agreement with what we saw in Section~\ref{sec:validation}, namely that GCMs generally produce higher velocities than those we predict here.

\subsection{Variation with Equilibrium Temperature}\label{sec:eqm_temp}

Finally, we model a range of hot Jupiters to determine the day and night side temperature contrast as a function of equilibrium temperature. Previous works have proposed that the highest temperature planets should have the poorest heat recirculation \citep[e.g.][]{cowan2011, perna2012}.
Intuitively, this is because as the equilibrium temperature increases, the wind speed increases and thus the wind crossing time-scale decreases. However, because the vertical heat flux scales as $\sim T_{\rm eq}^4$, the thermal time-scale goes as $\sim T_{\rm eq}^{-3}$, and so decreases much more quickly than the wind-crossing time.
As a result the efficiency declines faster than the wind speed increases, so the net heat transfer declines.

To see if this intuition is reproduced by our model, we take a hot Jupiter around a Sun-like star, vary the equilibrium temperature between 1000-3000~K, and see how $\Delta T/T_\mathrm{day}$ varies. We include opacity from TiO for the ultra-hot Jupiters with equilibrium temperatures above 2000~K as it is expected to be gaseous \citep[e.g.][]{sharp2007}. We also include thermal dissociation of TiO, H$_2$O, H- and H$_2$, and H- opacity for these ultra-hot Jupiters, which have been shown to be important \citep[e.g.][]{arcangeli2018, parmentier2018}. The middle panel of Figure~\ref{fig:dt_t} shows $\Delta T/T_\mathrm{day}$ as a function of equilibrium temperature for various pressures. The left and right panels also show example temperature profiles for a subset of these models for the cooler (non-inverted) and hotter cases (with thermal inversions) respectively. We can clearly see two trends in both sets of models. Firstly, we confirm that $\Delta T/T_\mathrm{day}$ generally increases with equilibrium temperature as expected. The second trend we can see is that the temperature difference is greater for lower pressures. This is unsurprising given that the lower pressures have the lowest optical depth and lowest radiative timescales. Near 1~bar the difference is negligible for all equilibrium temperatures, but $\Delta T/T_\mathrm{day} \gtrsim 0.6$ is possible at P$\sim$0.01~bar with the strong incident radiation.

As TiO is introduced into the atmosphere at equilibrium temperatures $\geq2000$~K we see a significant change in the values of $\Delta T/T_\mathrm{day}$ (see middle panel Figure~\ref{fig:dt_t}). This step change in $\Delta T/T_\mathrm{day}$ is driven by the strong TiO optical opacity absorbing the stellar flux in the upper layers of the atmosphere and thus causing a thermal inversion. At this transition point of T$_\mathrm{eq}=2000$~K, we have run models both with and without TiO. For the model with TiO, $\Delta T/T_\mathrm{day}$ is greater for P$\lesssim0.03$~bar but lower for P$\gtrsim0.1$~bar than the model without TiO. This is caused by of the high optical opacity as well as a lack of significant infrared opacity \citep[see e.g.][]{molliere2015, gandhi2019}. The presence of TiO, which increases the optical opacity, and the thermal dissociation of H$_2$O, which decreases the infrared opacity, increase $\Delta T/T_\mathrm{day}$ for the lowest pressures by preventing the upper layers on the day side from cooling effectively (see Section~\ref{sec:wasp76}).

As we further increase T$_\mathrm{eq}$, $\Delta T/T_\mathrm{day}$ begins to decrease beyond T$_\mathrm{eq}\sim2500$~K. The decline is caused by the dissociation of H$_2$ on the day side, which we include for T$_\mathrm{eq} \geq 2000$~K. This releases thermal energy onto the night side by recombination \citep[e.g.][]{bell2018, tan2019}. This effect is important to include when modelling ultra-hot Jupiters because, without the inclusion of the dissociation energy, $\Delta T/T_\mathrm{day}$ would continue to increase. H$_2$ becomes most easily dissociated in the upper atmosphere due to the lower pressure and the higher temperature (see Figure~\ref{fig:dt_t}), so the turn over point in $\Delta T/T_\mathrm{day}$ occurs for lower T$_\mathrm{eq}$ for the lower pressures.

Similarly, note there is a slight up-tick in $\Delta T/T_\mathrm{day}$ for $\mathrm{P}=0.01$~bar and $\mathrm{P}=0.03$~bar at T$_\mathrm{eq}\sim2500$~K and T$_\mathrm{eq}\sim2600$~K respectively. This occurs because H$_2$O dissociation significantly reduces the infrared opacity, which increases the day side temperature (see right panel Figure~\ref{fig:dt_t}) and offsets the dissociation of H$_2$. As thermal dissociation occurs most strongly at lower pressures and higher temperatures, the up-tick occurs for $\mathrm{P}=0.01$~bar at a lower equilibrium temperature.

\subsubsection{Comparison to Observations}

Figure~\ref{fig:dt_t} shows the day-night temperature difference from observations of a number of hot and ultra-hot Jupiters. These were derived from \citet{keating2019} using observations for HD~189733~b \citep{knutson2007, knutson2009, knutson2012}, WASP-43~b \citep{stevenson2014, mendonca2018}, HD~209458~b \citep{crossfield2012, zellem2014}, CoRoT-2~b \citep{dang2018}, HD~149026~b \citep{zhang2018}, WASP-14~b \citep{wong2015_wasp14}, WASP-19~b \citep{wong2016}, HAT-P-7~b \citep{wong2016}, KELT-1~b \citep{beatty2019}, WASP-18~b \citep{maxted2013}, WASP-103~b \citep{kreidberg2018}, WASP-12~b \citep{cowan2012} and WASP-33~b \citep{zhang2018}. The redistribution of radiation is less efficient at higher equilibrium temperatures, resulting in larger day-night temperature differences and confirming the trend predicted by earlier theory \citep[e.g.][]{cowan2011, perna2012} as well as our wind model. $\Delta T/T_\mathrm{day}$ most closely matches our model at P$\sim0.01-0.1$~bar, consistent with where we expect the photosphere to be. The data also show a slight flattening/downward trend in $\Delta T/T_\mathrm{day}$ at high values of T$_\mathrm{eq}$, consistent with H$_2$ dissociation in our model.
 
In Figure~\ref{fig:td_tn} we show the day side and the night side temperatures from our model and the observations \citep{keating2019}.
We have modelled a cloud free atmosphere, but any cloud opacity will alter the optical depth and so change the pressure of the photosphere.
As a result we show several pressure values for $\Delta T/T_\mathrm{day}$ near to where we expect the photosphere to be.

Note that for the day side there is a shift in the temperature profiles at equilibrium temperatures T$_\mathrm{eq}\geq2000$~K as a result of the thermal inversion (see right panel Figure~\ref{fig:dt_t}) and therefore hotter temperatures occur at lower pressures. Both the day and the night side agree well with the observations, with the best fit to the data at pressures $\sim10^{-1}-10^{-2}$~bar. Whilst the day side temperature increases almost linearly with equilibrium temperature, the night side temperature only shows a significant increase at the very hottest temperatures when H$_2$ recombination deposits significant flux onto the night side.

\section{Conclusions}\label{sec:conclusion}

We have constructed a self-consistent day-night wind model for hot Jupiter atmospheres and incorporated it into the radiative-convective equilibrium atmospheric model GENESIS \citep{gandhi2017}, providing an intermediate bridge between 1-D radiative-convective equilibrium atmospheric models and more complex 3-D GCMs.

We used the wind model to explore the radiative and chemical equilibrium day and night side atmosphere of WASP-76~b, an ultra-hot Jupiter with an equilibrium temperature $\sim$2200~K. We chose this planet because it has shown Fe condensation on the night side \citep{ehrenreich2020} and a thermal inversion from day side HST and Spitzer observations \citep{fu2020}. We modelled its atmosphere with a thermal inversion by introducing TiO at solar abundance, which gave good agreement with the HST and Spitzer observations. Our wind model also shows that the night side is cooler than the Fe condensation curve for pressures $\lesssim0.1$~bar, consistent with the rainout of Fe seen by \citet{ehrenreich2020}.

We also modelled the atmosphere of WASP-43~b in radiative equilibrium, a planet with high precision phase curves \citep{stevenson2014, stevenson2017} which has previously been interpreted through the lens of GCMs \citep{kataria2015}. We found good agreement with the predicted atmospheric profile from GCMs. We also compared the temperature profiles from the wind model to the retrieved day and night side profiles with the HyDRA retrieval code~\citep{gandhi2018}. The retrieved and wind model values agree well in the upper atmosphere but we did see deviations at pressures $\gtrsim10^{-1}$~bar for the night side where the observations are not sensitive, potentially due to cloud formation. This is because the retrieval fixes an isotherm with a wide uncertainty in the absence of constraining data. However, our wind model predicts a strong temperature gradient on the night side below the observable photosphere because eventually the night side temperature must equal the day side at high pressure.
This highlights the importance of factoring in heat transport between the day and night sides in inferring atmospheric properties from observations.

Finally, we explored how the temperature difference ($\Delta T/T_\mathrm{day}$) between the day and night varies with equilibrium temperature and pressure. We modelled planets above T$_\mathrm{eq} = 2000$~K with TiO, expected to be gaseous at such temperatures, and thermal dissociation, which has been shown to be relevant for such ultra-hot planets \citep[e.g.][]{arcangeli2018, parmentier2018}. We found that for T$_\mathrm{eq} \lesssim 2500$~K, $\Delta T/T_\mathrm{day}$ increases with increasing equilibrium temperature because the thermal time-scale of the atmosphere declines rapidly with increasing $T_{\rm eq}$, reducing the efficiency of winds at transporting heat.
At equilibrium temperatures in excess of $\sim$2500~K, the dissociation of H$_2$ deposits a significant amount of energy onto the night side by its recombination. This acts against the trend in $\Delta T/T_\mathrm{day}$ seen for lower temperatures and reduces the temperature difference between the day and night side. Our result is consistent with previous work \citep[e.g.][]{cowan2011, perna2012, komacek2016, bell2018} as well as observations of a number of hot Jupiters \citep{keating2019}. 

In the future our model could be extended include heat transport in rocky planets \citep[e.g.][]{koll2016, wordsworth2015}, and could be used to predict the strength of related effects such as thermal tides \citep{arras2010, lee2019}. Another improvement would be to extend the model to include more sides than just the two hemispheres, which would allow for a more realistic prescription for incident radiation as a cosine of the longitude. There is also potential to include the wind flux into retrievals of high resolution spectra \citep{brogi2017, brogi2019, gandhi2019_hr, gibson2020}, which have already shown some constraints on wind speeds \citep{snellen2010, brogi2016}. We may also use this model to explore non-irradiated objects and their variability \citep[e.g.][]{tan2020} by introducing cloudy and clear regions of the atmosphere which transfer flux between them. In addition, recent work has also shown the potential of 2-D retrievals of phase resolved spectra of hot Jupiters \citep[e.g.][]{irwin2020, feng2020}. Understanding how winds transport both flux and material across the two sides of the atmosphere can help determine how well mixing occurs in the atmosphere and place constraints on disequilibrium chemistry. This may also help us predict the night side atmosphere, particularly because that the cooler temperatures often induce greater disequilibrium and condensation of species. Given newly developed high resolution spectrographs (e.g. SPIRou, CARMENES and GIANO) and the impending arrival of space based facilities such as JWST and ARIEL, phase resolved measurements of exoplanets will only increase in both quantity and quality, so there is a real and growing need for tools to rapidly and accurately model large numbers of exoplanet atmospheres.

\section*{Acknowledgements}

SG acknowledges support from the UK Science and Technology Facilities Council (STFC) research grant ST/S000631/1. ASJ thanks the UK Marshall Aid Commemoration Commission for a scholarship which enabled this work. This research was supported in part by the National Science Foundation under Grant No. NSF PHY-1748958. The Flatiron Institute is supported by the Simons Foundation. We thank Nikku Madhusudhan and Adam Showman for helpful comments on this manuscript. We also thank Joanna Barstow and Ivan Hubeny for a careful review of our manuscript.

\section*{Data Availability}
The models underlying this article will be shared on reasonable request to the corresponding author.




\bibliographystyle{mnras}
\bibliography{refs} 

\begin{thebibliography}{}
\makeatletter
\relax
\def\mn@urlcharsother{\let\do\@makeother \do\$\do\&\do\#\do\^\do\_\do\%\do\~}
\def\mn@doi{\begingroup\mn@urlcharsother \@ifnextchar [ {\mn@doi@}
  {\mn@doi@[]}}
\def\mn@doi@[#1]#2{\def\@tempa{#1}\ifx\@tempa\@empty \href
  {http://dx.doi.org/#2} {doi:#2}\else \href {http://dx.doi.org/#2} {#1}\fi
  \endgroup}
\def\mn@eprint#1#2{\mn@eprint@#1:#2::\@nil}
\def\mn@eprint@arXiv#1{\href {http://arxiv.org/abs/#1} {{\tt arXiv:#1}}}
\def\mn@eprint@dblp#1{\href {http://dblp.uni-trier.de/rec/bibtex/#1.xml}
  {dblp:#1}}
\def\mn@eprint@#1:#2:#3:#4\@nil{\def\@tempa {#1}\def\@tempb {#2}\def\@tempc
  {#3}\ifx \@tempc \@empty \let \@tempc \@tempb \let \@tempb \@tempa \fi \ifx
  \@tempb \@empty \def\@tempb {arXiv}\fi \@ifundefined
  {mn@eprint@\@tempb}{\@tempb:\@tempc}{\expandafter \expandafter \csname
  mn@eprint@\@tempb\endcsname \expandafter{\@tempc}}}

\bibitem[\protect\citeauthoryear{{Arcangeli} et~al.,}{{Arcangeli}
  et~al.}{2018}]{arcangeli2018}
{Arcangeli} J.,  et~al., 2018, \mn@doi [\apjl] {10.3847/2041-8213/aab272},
  \href {https://ui.adsabs.harvard.edu/abs/2018ApJ...855L..30A} {855, L30}

\bibitem[\protect\citeauthoryear{{Arcangeli} et~al.,}{{Arcangeli}
  et~al.}{2019}]{arcangeli2019}
{Arcangeli} J.,  et~al., 2019, \mn@doi [\aap] {10.1051/0004-6361/201834891},
  \href {https://ui.adsabs.harvard.edu/abs/2019A&A...625A.136A} {625, A136}

\bibitem[\protect\citeauthoryear{{Arras} \& {Socrates}}{{Arras} \&
  {Socrates}}{2010}]{arras2010}
{Arras} P.,  {Socrates} A.,  2010, \mn@doi [\apj] {10.1088/0004-637X/714/1/1},
  \href {https://ui.adsabs.harvard.edu/abs/2010ApJ...714....1A} {714, 1}

\bibitem[\protect\citeauthoryear{{Asplund}, {Grevesse}, {Sauval}  \&
  {Scott}}{{Asplund} et~al.}{2009}]{asplund2009}
{Asplund} M.,  {Grevesse} N.,  {Sauval} A.~J.,   {Scott} P.,  2009, \mn@doi
  [\araa] {10.1146/annurev.astro.46.060407.145222}, \href
  {http://adsabs.harvard.edu/abs/2009ARA%26A..47..481A} {47, 481}

\bibitem[\protect\citeauthoryear{{Barber}, {Strange}, {Hill}, {Polyansky},
  {Mellau}, {Yurchenko}  \& {Tennyson}}{{Barber} et~al.}{2014}]{barber2014}
{Barber} R.~J.,  {Strange} J.~K.,  {Hill} C.,  {Polyansky} O.~L.,  {Mellau}
  G.~C.,  {Yurchenko} S.~N.,   {Tennyson} J.,  2014, \mn@doi [Mon. Not. R.
  Astron. Soc.] {10.1093/mnras/stt2011}, \href
  {http://adsabs.harvard.edu/abs/2014MNRAS.437.1828B} {437, 1828}

\bibitem[\protect\citeauthoryear{{Beatty}, {Marley}, {Gaudi}, {Col{\'o}n},
  {Fortney}  \& {Showman}}{{Beatty} et~al.}{2019}]{beatty2019}
{Beatty} T.~G.,  {Marley} M.~S.,  {Gaudi} B.~S.,  {Col{\'o}n} K.~D.,  {Fortney}
  J.~J.,   {Showman} A.~P.,  2019, \mn@doi [\aj] {10.3847/1538-3881/ab33fc},
  \href {https://ui.adsabs.harvard.edu/abs/2019AJ....158..166B} {158, 166}

\bibitem[\protect\citeauthoryear{{Bell} \& {Berrington}}{{Bell} \&
  {Berrington}}{1987}]{bell1987}
{Bell} K.~L.,  {Berrington} K.~A.,  1987, \mn@doi [Journal of Physics B Atomic
  Molecular Physics] {10.1088/0022-3700/20/4/019}, \href
  {https://ui.adsabs.harvard.edu/\#abs/1987JPhB...20..801B} {20, 801}

\bibitem[\protect\citeauthoryear{{Bell} \& {Cowan}}{{Bell} \&
  {Cowan}}{2018}]{bell2018}
{Bell} T.~J.,  {Cowan} N.~B.,  2018, \mn@doi [\apjl]
  {10.3847/2041-8213/aabcc8}, \href
  {https://ui.adsabs.harvard.edu/abs/2018ApJ...857L..20B} {857, L20}

\bibitem[\protect\citeauthoryear{{Brogi} \& {Line}}{{Brogi} \&
  {Line}}{2019}]{brogi2019}
{Brogi} M.,  {Line} M.~R.,  2019, \mn@doi [\aj] {10.3847/1538-3881/aaffd3},
  \href {https://ui.adsabs.harvard.edu/abs/2019AJ....157..114B} {157, 114}

\bibitem[\protect\citeauthoryear{{Brogi}, {de Kok}, {Albrecht}, {Snellen},
  {Birkby}  \& {Schwarz}}{{Brogi} et~al.}{2016}]{brogi2016}
{Brogi} M.,  {de Kok} R.~J.,  {Albrecht} S.,  {Snellen} I.~A.~G.,  {Birkby}
  J.~L.,   {Schwarz} H.,  2016, \mn@doi [\apj] {10.3847/0004-637X/817/2/106},
  \href {http://adsabs.harvard.edu/abs/2016ApJ...817..106B} {817, 106}

\bibitem[\protect\citeauthoryear{{Brogi}, {Line}, {Bean}, {D{\'e}sert}  \&
  {Schwarz}}{{Brogi} et~al.}{2017}]{brogi2017}
{Brogi} M.,  {Line} M.,  {Bean} J.,  {D{\'e}sert} J.~M.,   {Schwarz} H.,  2017,
  \mn@doi [\apjl] {10.3847/2041-8213/aa6933}, \href
  {https://ui.adsabs.harvard.edu/abs/2017ApJ...839L...2B} {839, L2}

\bibitem[\protect\citeauthoryear{{Burrows}, {Budaj}  \& {Hubeny}}{{Burrows}
  et~al.}{2008}]{burrows2008}
{Burrows} A.,  {Budaj} J.,   {Hubeny} I.,  2008, \mn@doi [\apj]
  {10.1086/533518}, \href {http://adsabs.harvard.edu/abs/2008ApJ...678.1436B}
  {678, 1436}

\bibitem[\protect\citeauthoryear{{Castelli} \& {Kurucz}}{{Castelli} \&
  {Kurucz}}{2003}]{kurucz_model}
{Castelli} F.,  {Kurucz} R.~L.,  2003, Symposium - International Astronomical
  Union, \href {https://ui.adsabs.harvard.edu/abs/2003IAUS..210P.A20C} {210,
  A20}

\bibitem[\protect\citeauthoryear{{Chemke} \& {Kaspi}}{{Chemke} \&
  {Kaspi}}{2015}]{2015JAtS...72.3891C}
{Chemke} R.,  {Kaspi} Y.,  2015, \mn@doi [Journal of Atmospheric Sciences]
  {10.1175/JAS-D-15-0007.1}, \href
  {http://adsabs.harvard.edu/abs/2015JAtS...72.3891C} {72, 3891}

\bibitem[\protect\citeauthoryear{{Chubb}, {Tennyson}  \& {Yurchenko}}{{Chubb}
  et~al.}{2020}]{chubb2020}
{Chubb} K.~L.,  {Tennyson} J.,   {Yurchenko} S.~N.,  2020, \mn@doi [\mnras]
  {10.1093/mnras/staa229}, \href
  {https://ui.adsabs.harvard.edu/abs/2020MNRAS.493.1531C} {493, 1531}

\bibitem[\protect\citeauthoryear{Coles, Yurchenko  \& Tennyson}{Coles
  et~al.}{2019}]{coles2019}
Coles P.~A.,  Yurchenko S.~N.,   Tennyson J.,  2019, \mn@doi [\mnras]
  {10.1093/mnras/stz2778}, 490, 4638

\bibitem[\protect\citeauthoryear{{Cooper} \& {Showman}}{{Cooper} \&
  {Showman}}{2006}]{cooper2006}
{Cooper} C.~S.,  {Showman} A.~P.,  2006, \mn@doi [\apj] {10.1086/506312}, \href
  {https://ui.adsabs.harvard.edu/abs/2006ApJ...649.1048C} {649, 1048}

\bibitem[\protect\citeauthoryear{{Cowan} \& {Agol}}{{Cowan} \&
  {Agol}}{2008}]{cowan2008}
{Cowan} N.~B.,  {Agol} E.,  2008, \mn@doi [\apjl] {10.1086/588553}, \href
  {http://adsabs.harvard.edu/abs/2008ApJ...678L.129C} {678, L129}

\bibitem[\protect\citeauthoryear{{Cowan} \& {Agol}}{{Cowan} \&
  {Agol}}{2011}]{cowan2011}
{Cowan} N.~B.,  {Agol} E.,  2011, \mn@doi [\apj] {10.1088/0004-637X/729/1/54},
  \href {http://adsabs.harvard.edu/abs/2011ApJ...729...54C} {729, 54}

\bibitem[\protect\citeauthoryear{{Cowan}, {Machalek}, {Croll}, {Shekhtman},
  {Burrows}, {Deming}, {Greene}  \& {Hora}}{{Cowan} et~al.}{2012}]{cowan2012}
{Cowan} N.~B.,  {Machalek} P.,  {Croll} B.,  {Shekhtman} L.~M.,  {Burrows} A.,
  {Deming} D.,  {Greene} T.,   {Hora} J.~L.,  2012, \mn@doi [\apj]
  {10.1088/0004-637X/747/1/82}, \href
  {https://ui.adsabs.harvard.edu/abs/2012ApJ...747...82C} {747, 82}

\bibitem[\protect\citeauthoryear{{Crossfield}, {Knutson}, {Fortney}, {Showman},
  {Cowan}  \& {Deming}}{{Crossfield} et~al.}{2012}]{crossfield2012}
{Crossfield} I. J.~M.,  {Knutson} H.,  {Fortney} J.,  {Showman} A.~P.,  {Cowan}
  N.~B.,   {Deming} D.,  2012, \mn@doi [\apj] {10.1088/0004-637X/752/2/81},
  \href {https://ui.adsabs.harvard.edu/abs/2012ApJ...752...81C} {752, 81}

\bibitem[\protect\citeauthoryear{{Dang} et~al.,}{{Dang}
  et~al.}{2018}]{dang2018}
{Dang} L.,  et~al., 2018, \mn@doi [Nature Astronomy]
  {10.1038/s41550-017-0351-6}, \href
  {https://ui.adsabs.harvard.edu/abs/2018NatAs...2..220D} {2, 220}

\bibitem[\protect\citeauthoryear{{Dobbs-Dixon} \& {Agol}}{{Dobbs-Dixon} \&
  {Agol}}{2013}]{dobbs-dixon2013}
{Dobbs-Dixon} I.,  {Agol} E.,  2013, \mn@doi [\mnras] {10.1093/mnras/stt1509},
  \href {https://ui.adsabs.harvard.edu/abs/2013MNRAS.435.3159D} {435, 3159}

\bibitem[\protect\citeauthoryear{{Drummond} et~al.,}{{Drummond}
  et~al.}{2020}]{drummond2020}
{Drummond} B.,  et~al., 2020, \mn@doi [\aap] {10.1051/0004-6361/201937153},
  \href {https://ui.adsabs.harvard.edu/abs/2020A&A...636A..68D} {636, A68}

\bibitem[\protect\citeauthoryear{{Ehrenreich} et~al.,}{{Ehrenreich}
  et~al.}{2020}]{ehrenreich2020}
{Ehrenreich} D.,  et~al., 2020, \mn@doi [\nat] {10.1038/s41586-020-2107-1},
  \href {https://ui.adsabs.harvard.edu/abs/2020Natur.580..597E} {580, 597}

\bibitem[\protect\citeauthoryear{{Feng}, {Line}  \& {Fortney}}{{Feng}
  et~al.}{2020}]{feng2020}
{Feng} Y.~K.,  {Line} M.~R.,   {Fortney} J.~J.,  2020, \mn@doi [\aj]
  {10.3847/1538-3881/aba8f9}, \href
  {https://ui.adsabs.harvard.edu/abs/2020AJ....160..137F} {160, 137}

\bibitem[\protect\citeauthoryear{{Flowers}, {Brogi}, {Rauscher}, {Kempton}  \&
  {Chiavassa}}{{Flowers} et~al.}{2019}]{flowers2019}
{Flowers} E.,  {Brogi} M.,  {Rauscher} E.,  {Kempton} E. M.~R.,   {Chiavassa}
  A.,  2019, \mn@doi [\aj] {10.3847/1538-3881/ab164c}, \href
  {https://ui.adsabs.harvard.edu/abs/2019AJ....157..209F} {157, 209}

\bibitem[\protect\citeauthoryear{{Fortney}, {Lodders}, {Marley}  \&
  {Freedman}}{{Fortney} et~al.}{2008}]{fortney2008}
{Fortney} J.~J.,  {Lodders} K.,  {Marley} M.~S.,   {Freedman} R.~S.,  2008,
  \mn@doi [\apj] {10.1086/528370}, \href
  {http://adsabs.harvard.edu/abs/2008ApJ...678.1419F} {678, 1419}

\bibitem[\protect\citeauthoryear{{Fu} et~al.,}{{Fu} et~al.}{2020}]{fu2020}
{Fu} G.,  et~al., 2020, arXiv e-prints, \href
  {https://ui.adsabs.harvard.edu/abs/2020arXiv200502568F} {p. arXiv:2005.02568}

\bibitem[\protect\citeauthoryear{{Gandhi} \& {Madhusudhan}}{{Gandhi} \&
  {Madhusudhan}}{2017}]{gandhi2017}
{Gandhi} S.,  {Madhusudhan} N.,  2017, \mn@doi [\mnras]
  {10.1093/mnras/stx1601}, \href
  {http://adsabs.harvard.edu/abs/2017MNRAS.472.2334G} {472, 2334}

\bibitem[\protect\citeauthoryear{{Gandhi} \& {Madhusudhan}}{{Gandhi} \&
  {Madhusudhan}}{2018}]{gandhi2018}
{Gandhi} S.,  {Madhusudhan} N.,  2018, \mn@doi [\mnras]
  {10.1093/mnras/stx2748}, \href
  {http://adsabs.harvard.edu/abs/2018MNRAS.474..271G} {474, 271}

\bibitem[\protect\citeauthoryear{{Gandhi} \& {Madhusudhan}}{{Gandhi} \&
  {Madhusudhan}}{2019}]{gandhi2019}
{Gandhi} S.,  {Madhusudhan} N.,  2019, \mn@doi [\mnras] {10.1093/mnras/stz751},
  \href {https://ui.adsabs.harvard.edu/abs/2019MNRAS.485.5817G} {485, 5817}

\bibitem[\protect\citeauthoryear{{Gandhi}, {Madhusudhan}, {Hawker}  \&
  {Piette}}{{Gandhi} et~al.}{2019}]{gandhi2019_hr}
{Gandhi} S.,  {Madhusudhan} N.,  {Hawker} G.,   {Piette} A.,  2019, \mn@doi
  [\aj] {10.3847/1538-3881/ab4efc}, \href
  {https://ui.adsabs.harvard.edu/abs/2019AJ....158..228G} {158, 228}

\bibitem[\protect\citeauthoryear{{Gandhi}, {Madhusudhan}  \&
  {Mandell}}{{Gandhi} et~al.}{2020a}]{gandhi2020_h-}
{Gandhi} S.,  {Madhusudhan} N.,   {Mandell} A.,  2020a, \mn@doi [\aj]
  {10.3847/1538-3881/ab845e}, \href
  {https://ui.adsabs.harvard.edu/abs/2020AJ....159..232G} {159, 232}

\bibitem[\protect\citeauthoryear{{Gandhi} et~al.,}{{Gandhi}
  et~al.}{2020b}]{gandhi2020_cs}
{Gandhi} S.,  et~al., 2020b, \mn@doi [\mnras] {10.1093/mnras/staa981}, \href
  {https://ui.adsabs.harvard.edu/abs/2020MNRAS.495..224G} {495, 224}

\bibitem[\protect\citeauthoryear{Garaud, Gagnier  \& Verhoeven}{Garaud
  et~al.}{2017}]{0004-637X-837-2-133}
Garaud P.,  Gagnier D.,   Verhoeven J.,  2017, The Astrophysical Journal, 837,
  133

\bibitem[\protect\citeauthoryear{{Gibson} et~al.,}{{Gibson}
  et~al.}{2020}]{gibson2020}
{Gibson} N.~P.,  et~al., 2020, \mn@doi [\mnras] {10.1093/mnras/staa228}, \href
  {https://ui.adsabs.harvard.edu/abs/2020MNRAS.493.2215G} {493, 2215}

\bibitem[\protect\citeauthoryear{Gough}{Gough}{2012}]{Gough2012}
Gough D.~O.,  2012, \mn@doi [{ISRN} Astronomy and Astrophysics]
  {10.5402/2012/987275}, 2012, 1

\bibitem[\protect\citeauthoryear{{Hammond} \& {Pierrehumbert}}{{Hammond} \&
  {Pierrehumbert}}{2018}]{hammond2018}
{Hammond} M.,  {Pierrehumbert} R.~T.,  2018, \mn@doi [\apj]
  {10.3847/1538-4357/aaec03}, \href
  {https://ui.adsabs.harvard.edu/abs/2018ApJ...869...65H} {869, 65}

\bibitem[\protect\citeauthoryear{{Hargreaves}, {Gordon}, {Rey}, {Nikitin},
  {Tyuterev}, {Kochanov}  \& {Rothman}}{{Hargreaves}
  et~al.}{2020}]{hargreaves2020}
{Hargreaves} R.~J.,  {Gordon} I.~E.,  {Rey} M.,  {Nikitin} A.~V.,  {Tyuterev}
  V.~G.,  {Kochanov} R.~V.,   {Rothman} L.~S.,  2020, \mn@doi [\apjs]
  {10.3847/1538-4365/ab7a1a}, \href
  {https://ui.adsabs.harvard.edu/abs/2020ApJS..247...55H} {247, 55}

\bibitem[\protect\citeauthoryear{{Haynes}, {Mandell}, {Madhusudhan}, {Deming}
  \& {Knutson}}{{Haynes} et~al.}{2015}]{haynes2015}
{Haynes} K.,  {Mandell} A.~M.,  {Madhusudhan} N.,  {Deming} D.,   {Knutson} H.,
   2015, \mn@doi [\apj] {10.1088/0004-637X/806/2/146}, \href
  {http://adsabs.harvard.edu/abs/2015ApJ...806..146H} {806, 146}

\bibitem[\protect\citeauthoryear{{Heng} \& {Tsai}}{{Heng} \&
  {Tsai}}{2016}]{heng2016}
{Heng} K.,  {Tsai} S.-M.,  2016, \mn@doi [\apj] {10.3847/0004-637X/829/2/104},
  \href {http://adsabs.harvard.edu/abs/2016ApJ...829..104H} {829, 104}

\bibitem[\protect\citeauthoryear{Hrebtov, Ilyushin  \& Krasinsky}{Hrebtov
  et~al.}{2010}]{PhysRevE.81.016315}
Hrebtov M.~Y.,  Ilyushin B.~B.,   Krasinsky D.~V.,  2010, \mn@doi [Phys. Rev.
  E] {10.1103/PhysRevE.81.016315}, 81, 016315

\bibitem[\protect\citeauthoryear{{Hubeny}}{{Hubeny}}{2017}]{hubeny2017}
{Hubeny} I.,  2017, \mn@doi [\mnras] {10.1093/mnras/stx758}, \href
  {http://adsabs.harvard.edu/abs/2017MNRAS.469..841H} {469, 841}

\bibitem[\protect\citeauthoryear{{Hubeny} \& {Mihalas}}{{Hubeny} \&
  {Mihalas}}{2014}]{hubenybook}
{Hubeny} I.,  {Mihalas} D.,  2014, {Theory of Stellar Atmospheres}.
Princeton University Press

\bibitem[\protect\citeauthoryear{{Hubeny}, {Burrows}  \& {Sudarsky}}{{Hubeny}
  et~al.}{2003}]{hubeny2003}
{Hubeny} I.,  {Burrows} A.,   {Sudarsky} D.,  2003, \mn@doi [\apj]
  {10.1086/377080}, \href {http://adsabs.harvard.edu/abs/2003ApJ...594.1011H}
  {594, 1011}

\bibitem[\protect\citeauthoryear{{Irwin}, {Parmentier}, {Taylor}, {Barstow},
  {Aigrain}, {Lee}  \& {Garland }}{{Irwin} et~al.}{2020}]{irwin2020}
{Irwin} P. G.~J.,  {Parmentier} V.,  {Taylor} J.,  {Barstow} J.,  {Aigrain} S.,
   {Lee} G. K.~H.,   {Garland } R.,  2020, \mn@doi [\mnras]
  {10.1093/mnras/staa238}, \href
  {https://ui.adsabs.harvard.edu/abs/2020MNRAS.493..106I} {493, 106}

\bibitem[\protect\citeauthoryear{Jermyn}{Jermyn}{2015}]{https://doi.org/10.7907/z90z716m}
Jermyn A.~S.,  2015, The Atmospheric Dynamics of Pulsar Companions,
  \mn@doi{10.7907/z90z716m}, \url {https://thesis.library.caltech.edu/9019/}

\bibitem[\protect\citeauthoryear{Jermyn, Tout  \& Ogilvie}{Jermyn
  et~al.}{2017}]{doi:10.1093/mnras/stx831}
Jermyn A.~S.,  Tout C.~A.,   Ogilvie G.~I.,  2017, \mn@doi [Monthly Notices of
  the Royal Astronomical Society] {10.1093/mnras/stx831}, 469, 1768

\bibitem[\protect\citeauthoryear{Jermyn, Lesaffre, Tout  \& Chitre}{Jermyn
  et~al.}{2018}]{doi:10.1093/mnras/sty255}
Jermyn A.~S.,  Lesaffre P.,  Tout C.~A.,   Chitre S.~M.,  2018, \mn@doi
  [Monthly Notices of the Royal Astronomical Society] {10.1093/mnras/sty255},
  476, 646

\bibitem[\protect\citeauthoryear{{John}}{{John}}{1988}]{john1988}
{John} T.~L.,  1988, \aap, \href
  {https://ui.adsabs.harvard.edu/\#abs/1988A&A...193..189J} {193, 189}

\bibitem[\protect\citeauthoryear{{Kataria}, {Showman}, {Fortney}, {Stevenson},
  {Line}, {Kreidberg}, {Bean}  \& {D{\'e}sert}}{{Kataria}
  et~al.}{2015}]{kataria2015}
{Kataria} T.,  {Showman} A.~P.,  {Fortney} J.~J.,  {Stevenson} K.~B.,  {Line}
  M.~R.,  {Kreidberg} L.,  {Bean} J.~L.,   {D{\'e}sert} J.-M.,  2015, \mn@doi
  [\apj] {10.1088/0004-637X/801/2/86}, \href
  {http://adsabs.harvard.edu/abs/2015ApJ...801...86K} {801, 86}

\bibitem[\protect\citeauthoryear{{Kataria}, {Sing}, {Lewis}, {Visscher},
  {Showman}, {Fortney}  \& {Marley}}{{Kataria} et~al.}{2016}]{kataria2016}
{Kataria} T.,  {Sing} D.~K.,  {Lewis} N.~K.,  {Visscher} C.,  {Showman} A.~P.,
  {Fortney} J.~J.,   {Marley} M.~S.,  2016, \mn@doi [\apj]
  {10.3847/0004-637X/821/1/9}, \href
  {http://adsabs.harvard.edu/abs/2016ApJ...821....9K} {821, 9}

\bibitem[\protect\citeauthoryear{{Keating}, {Cowan}  \& {Dang}}{{Keating}
  et~al.}{2019}]{keating2019}
{Keating} D.,  {Cowan} N.~B.,   {Dang} L.,  2019, \mn@doi [Nature Astronomy]
  {10.1038/s41550-019-0859-z}, \href
  {https://ui.adsabs.harvard.edu/abs/2019NatAs...3.1092K} {3, 1092}

\bibitem[\protect\citeauthoryear{{Kichatinov} \& {Rudiger}}{{Kichatinov} \&
  {Rudiger}}{1993}]{1993A&A...276...96K}
{Kichatinov} L.~L.,  {Rudiger} G.,  1993, \aap, \href
  {http://adsabs.harvard.edu/abs/1993A%26A...276...96K} {276, 96}

\bibitem[\protect\citeauthoryear{{Kippenhahn}, {Weigert}  \&
  {Weiss}}{{Kippenhahn} et~al.}{2012}]{kippenhahn2012}
{Kippenhahn} R.,  {Weigert} A.,   {Weiss} A.,  2012, {Stellar Structure and
  Evolution}.
Springer, \mn@doi{10.1007/978-3-642-30304-3}

\bibitem[\protect\citeauthoryear{{Knutson} et~al.,}{{Knutson}
  et~al.}{2007}]{knutson2007}
{Knutson} H.~A.,  et~al., 2007, \mn@doi [\nat] {10.1038/nature05782}, \href
  {http://adsabs.harvard.edu/abs/2007Natur.447..183K} {447, 183}

\bibitem[\protect\citeauthoryear{{Knutson} et~al.,}{{Knutson}
  et~al.}{2009a}]{knutson2009}
{Knutson} H.~A.,  et~al., 2009a, \mn@doi [\apj] {10.1088/0004-637X/690/1/822},
  \href {http://adsabs.harvard.edu/abs/2009ApJ...690..822K} {690, 822}

\bibitem[\protect\citeauthoryear{{Knutson}, {Charbonneau}, {Cowan}, {Fortney},
  {Showman}, {Agol}  \& {Henry}}{{Knutson} et~al.}{2009b}]{knutson2009b}
{Knutson} H.~A.,  {Charbonneau} D.,  {Cowan} N.~B.,  {Fortney} J.~J.,
  {Showman} A.~P.,  {Agol} E.,   {Henry} G.~W.,  2009b, \mn@doi [\apj]
  {10.1088/0004-637X/703/1/769}, \href
  {http://adsabs.harvard.edu/abs/2009ApJ...703..769K} {703, 769}

\bibitem[\protect\citeauthoryear{{Knutson} et~al.,}{{Knutson}
  et~al.}{2012}]{knutson2012}
{Knutson} H.~A.,  et~al., 2012, \mn@doi [\apj] {10.1088/0004-637X/754/1/22},
  \href {http://adsabs.harvard.edu/abs/2012ApJ...754...22K} {754, 22}

\bibitem[\protect\citeauthoryear{{Koll} \& {Abbot}}{{Koll} \&
  {Abbot}}{2016}]{koll2016}
{Koll} D. D.~B.,  {Abbot} D.~S.,  2016, \mn@doi [\apj]
  {10.3847/0004-637X/825/2/99}, \href
  {https://ui.adsabs.harvard.edu/abs/2016ApJ...825...99K} {825, 99}

\bibitem[\protect\citeauthoryear{{Komacek} \& {Showman}}{{Komacek} \&
  {Showman}}{2016}]{komacek2016}
{Komacek} T.~D.,  {Showman} A.~P.,  2016, \mn@doi [\apj]
  {10.3847/0004-637X/821/1/16}, \href
  {https://ui.adsabs.harvard.edu/abs/2016ApJ...821...16K} {821, 16}

\bibitem[\protect\citeauthoryear{{Komacek} \& {Tan}}{{Komacek} \&
  {Tan}}{2018}]{komacek2018}
{Komacek} T.~D.,  {Tan} X.,  2018, \mn@doi [Research Notes of the American
  Astronomical Society] {10.3847/2515-5172/aac5e7}, \href
  {https://ui.adsabs.harvard.edu/abs/2018RNAAS...2...36K} {2, 36}

\bibitem[\protect\citeauthoryear{{Komacek}, {Showman}  \& {Tan}}{{Komacek}
  et~al.}{2017}]{komacek2017}
{Komacek} T.~D.,  {Showman} A.~P.,   {Tan} X.,  2017, \mn@doi [\apj]
  {10.3847/1538-4357/835/2/198}, \href
  {https://ui.adsabs.harvard.edu/abs/2017ApJ...835..198K} {835, 198}

\bibitem[\protect\citeauthoryear{{Kramida}, {Ralchenko}, {Nave}  \&
  {Reader}}{{Kramida} et~al.}{2018}]{kramida2018}
{Kramida} A.,  {Ralchenko} Y.,  {Nave} G.,   {Reader} J.,  2018, in APS
  Division of Atomic, Molecular and Optical Physics Meeting Abstracts. p.
  M01.004

\bibitem[\protect\citeauthoryear{{Kreidberg} et~al.,}{{Kreidberg}
  et~al.}{2018}]{kreidberg2018}
{Kreidberg} L.,  et~al., 2018, \mn@doi [\aj] {10.3847/1538-3881/aac3df}, \href
  {http://adsabs.harvard.edu/abs/2018AJ....156...17K} {156, 17}

\bibitem[\protect\citeauthoryear{{Kurucz}}{{Kurucz}}{1979}]{Kurucz_1979_paper}
{Kurucz} R.~L.,  1979, \mn@doi [\apjs] {10.1086/190589}, \href
  {http://adsabs.harvard.edu/abs/1979ApJS...40....1K} {40, 1}

\bibitem[\protect\citeauthoryear{{Lee} \& {Murakami}}{{Lee} \&
  {Murakami}}{2019}]{lee2019}
{Lee} U.,  {Murakami} D.,  2019, \mn@doi [\mnras] {10.1093/mnras/stz1822},
  \href {https://ui.adsabs.harvard.edu/abs/2019MNRAS.488.1960L} {488, 1960}

\bibitem[\protect\citeauthoryear{Li, Gordon, Rothman, Tan, Hu, Kassi, Campargue
   \& Medvedev}{Li et~al.}{2015}]{li2015}
Li G.,  Gordon I.~E.,  Rothman L.~S.,  Tan Y.,  Hu S.-M.,  Kassi S.,  Campargue
  A.,   Medvedev E.~S.,  2015, \mn@doi [The Astrophysical Journal Supplement
  Series] {10.1088/0067-0049/216/1/15}, 216, 15

\bibitem[\protect\citeauthoryear{Liu \& Schneider}{Liu \&
  Schneider}{2010}]{Liu2010}
Liu J.,  Schneider T.,  2010, \mn@doi [Journal of the Atmospheric Sciences]
  {10.1175/2010jas3492.1}, 67, 3652

\bibitem[\protect\citeauthoryear{{Lothringer}, {Barman}  \&
  {Koskinen}}{{Lothringer} et~al.}{2018}]{lothringer2018}
{Lothringer} J.~D.,  {Barman} T.,   {Koskinen} T.,  2018, \mn@doi [\apj]
  {10.3847/1538-4357/aadd9e}, \href
  {https://ui.adsabs.harvard.edu/abs/2018ApJ...866...27L} {866, 27}

\bibitem[\protect\citeauthoryear{{Maxted} et~al.,}{{Maxted}
  et~al.}{2013}]{maxted2013}
{Maxted} P.~F.~L.,  et~al., 2013, \mn@doi [\mnras] {10.1093/mnras/sts231},
  \href {https://ui.adsabs.harvard.edu/abs/2013MNRAS.428.2645M} {428, 2645}

\bibitem[\protect\citeauthoryear{{Mayne} et~al.,}{{Mayne}
  et~al.}{2014}]{mayne2014}
{Mayne} N.~J.,  et~al., 2014, \mn@doi [\aap] {10.1051/0004-6361/201322174},
  \href {https://ui.adsabs.harvard.edu/abs/2014A&A...561A...1M} {561, A1}

\bibitem[\protect\citeauthoryear{{McKemmish}, {Masseron}, {Hoeijmakers},
  {P{\'e}rez-Mesa}, {Grimm}, {Yurchenko}  \& {Tennyson}}{{McKemmish}
  et~al.}{2019}]{mckemmish2019}
{McKemmish} L.~K.,  {Masseron} T.,  {Hoeijmakers} H.~J.,  {P{\'e}rez-Mesa} V.,
  {Grimm} S.~L.,  {Yurchenko} S.~N.,   {Tennyson} J.,  2019, \mn@doi [\mnras]
  {10.1093/mnras/stz1818}, \href
  {https://ui.adsabs.harvard.edu/abs/2019MNRAS.488.2836M} {488, 2836}

\bibitem[\protect\citeauthoryear{{Mendon{\c{c}}a}, {Malik}, {Demory}  \&
  {Heng}}{{Mendon{\c{c}}a} et~al.}{2018}]{mendonca2018}
{Mendon{\c{c}}a} J.~M.,  {Malik} M.,  {Demory} B.-O.,   {Heng} K.,  2018,
  \mn@doi [\aj] {10.3847/1538-3881/aaaebc}, \href
  {https://ui.adsabs.harvard.edu/abs/2018AJ....155..150M} {155, 150}

\bibitem[\protect\citeauthoryear{{Mikal-Evans}, {Sing}, {Kataria}, {Wakeford},
  {Mayne}, {Lewis}, {Barstow}  \& {Spake}}{{Mikal-Evans}
  et~al.}{2020}]{mikal-evans2020}
{Mikal-Evans} T.,  {Sing} D.~K.,  {Kataria} T.,  {Wakeford} H.~R.,  {Mayne}
  N.~J.,  {Lewis} N.~K.,  {Barstow} J.~K.,   {Spake} J.~J.,  2020, \mn@doi
  [\mnras] {10.1093/mnras/staa1628}, \href
  {https://ui.adsabs.harvard.edu/abs/2020MNRAS.496.1638M} {496, 1638}

\bibitem[\protect\citeauthoryear{{Molli{\`e}re}, {van Boekel}, {Dullemond},
  {Henning}  \& {Mordasini}}{{Molli{\`e}re} et~al.}{2015}]{molliere2015}
{Molli{\`e}re} P.,  {van Boekel} R.,  {Dullemond} C.,  {Henning} T.,
  {Mordasini} C.,  2015, \mn@doi [\apj] {10.1088/0004-637X/813/1/47}, \href
  {http://adsabs.harvard.edu/abs/2015ApJ...813...47M} {813, 47}

\bibitem[\protect\citeauthoryear{{Parmentier} et~al.,}{{Parmentier}
  et~al.}{2018}]{parmentier2018}
{Parmentier} V.,  et~al., 2018, \mn@doi [\aap] {10.1051/0004-6361/201833059},
  \href {https://ui.adsabs.harvard.edu/abs/2018A&A...617A.110P} {617, A110}

\bibitem[\protect\citeauthoryear{{Perna}, {Heng}  \& {Pont}}{{Perna}
  et~al.}{2012}]{perna2012}
{Perna} R.,  {Heng} K.,   {Pont} F.,  2012, \mn@doi [\apj]
  {10.1088/0004-637X/751/1/59}, \href
  {https://ui.adsabs.harvard.edu/abs/2012ApJ...751...59P} {751, 59}

\bibitem[\protect\citeauthoryear{{Piette}, {Madhusudhan}, {McKemmish},
  {Gandhi}, {Masseron}  \& {Welbanks}}{{Piette} et~al.}{2020}]{piette2020}
{Piette} A. A.~A.,  {Madhusudhan} N.,  {McKemmish} L.~K.,  {Gandhi} S.,
  {Masseron} T.,   {Welbanks} L.,  2020, \mn@doi [\mnras]
  {10.1093/mnras/staa1592}, \href
  {https://ui.adsabs.harvard.edu/abs/2020MNRAS.496.3870P} {496, 3870}

\bibitem[\protect\citeauthoryear{{Polyansky}, {Kyuberis}, {Zobov}, {Tennyson},
  {Yurchenko}  \& {Lodi}}{{Polyansky} et~al.}{2018}]{polyansky2018}
{Polyansky} O.~L.,  {Kyuberis} A.~A.,  {Zobov} N.~F.,  {Tennyson} J.,
  {Yurchenko} S.~N.,   {Lodi} L.,  2018, \mn@doi [\mnras]
  {10.1093/mnras/sty1877}, \href
  {http://adsabs.harvard.edu/abs/2018MNRAS.480.2597P} {480, 2597}

\bibitem[\protect\citeauthoryear{{Rauscher} \& {Menou}}{{Rauscher} \&
  {Menou}}{2010}]{rauscher2010}
{Rauscher} E.,  {Menou} K.,  2010, \mn@doi [\apj]
  {10.1088/0004-637X/714/2/1334}, \href
  {https://ui.adsabs.harvard.edu/abs/2010ApJ...714.1334R} {714, 1334}

\bibitem[\protect\citeauthoryear{{Rauscher} \& {Menou}}{{Rauscher} \&
  {Menou}}{2013}]{rauscher2013}
{Rauscher} E.,  {Menou} K.,  2013, \mn@doi [\apj]
  {10.1088/0004-637X/764/1/103}, \href
  {https://ui.adsabs.harvard.edu/abs/2013ApJ...764..103R} {764, 103}

\bibitem[\protect\citeauthoryear{{Rhines}}{{Rhines}}{1975}]{1975JFM....69..417R}
{Rhines} P.~B.,  1975, \mn@doi [Journal of Fluid Mechanics]
  {10.1017/S0022112075001504}, \href
  {http://adsabs.harvard.edu/abs/1975JFM....69..417R} {69, 417}

\bibitem[\protect\citeauthoryear{{Richard} et~al.,}{{Richard}
  et~al.}{2012}]{richard2012}
{Richard} C.,  et~al., 2012, \mn@doi [\jqsrt] {10.1016/j.jqsrt.2011.11.004},
  \href {https://ui.adsabs.harvard.edu/abs/2012JQSRT.113.1276R} {113, 1276}

\bibitem[\protect\citeauthoryear{{Rothman} et~al.,}{{Rothman}
  et~al.}{2010}]{rothman2010}
{Rothman} L.~S.,  et~al., 2010, \mn@doi [JQSRT] {10.1016/j.jqsrt.2010.05.001},
  \href {http://adsabs.harvard.edu/abs/2010JQSRT.111.2139R} {111, 2139}

\bibitem[\protect\citeauthoryear{{R{\"u}diger}, {K{\"u}ker}  \&
  {Tereshin}}{{R{\"u}diger} et~al.}{2014}]{2014A&A...572L...7R}
{R{\"u}diger} G.,  {K{\"u}ker} M.,   {Tereshin} I.,  2014, \mn@doi [\aap]
  {10.1051/0004-6361/201424953}, \href
  {https://ui.adsabs.harvard.edu/#abs/2014A&A...572L...7R} {572, L7}

\bibitem[\protect\citeauthoryear{Scott \& Dunkerton}{Scott \&
  Dunkerton}{2004}]{doi:10.1002/2017GL072628}
Scott R.~K.,  Dunkerton T.~J.,  2004, \mn@doi [Geophysical Research Letters]
  {10.1002/2017GL072628}, 44, 3073

\bibitem[\protect\citeauthoryear{{Sharp} \& {Burrows}}{{Sharp} \&
  {Burrows}}{2007}]{sharp2007}
{Sharp} C.~M.,  {Burrows} A.,  2007, \mn@doi [\apjs] {10.1086/508708}, \href
  {https://ui.adsabs.harvard.edu/abs/2007ApJS..168..140S} {168, 140}

\bibitem[\protect\citeauthoryear{{Sheppard}, {Mandell}, {Tamburo}, {Gandhi},
  {Pinhas}, {Madhusudhan}  \& {Deming}}{{Sheppard} et~al.}{2017}]{sheppard2017}
{Sheppard} K.~B.,  {Mandell} A.~M.,  {Tamburo} P.,  {Gandhi} S.,  {Pinhas} A.,
  {Madhusudhan} N.,   {Deming} D.,  2017, \mn@doi [\apjl]
  {10.3847/2041-8213/aa9ae9}, \href
  {http://adsabs.harvard.edu/abs/2017ApJ...850L..32S} {850, L32}

\bibitem[\protect\citeauthoryear{{Showman} \& {Polvani}}{{Showman} \&
  {Polvani}}{2011}]{showman2011}
{Showman} A.~P.,  {Polvani} L.~M.,  2011, \mn@doi [\apj]
  {10.1088/0004-637X/738/1/71}, \href
  {https://ui.adsabs.harvard.edu/abs/2011ApJ...738...71S} {738, 71}

\bibitem[\protect\citeauthoryear{{Showman}, {Cooper}, {Fortney}  \&
  {Marley}}{{Showman} et~al.}{2008}]{showman2008}
{Showman} A.~P.,  {Cooper} C.~S.,  {Fortney} J.~J.,   {Marley} M.~S.,  2008,
  \mn@doi [\apj] {10.1086/589325}, \href
  {http://adsabs.harvard.edu/abs/2008ApJ...682..559S} {682, 559}

\bibitem[\protect\citeauthoryear{{Showman}, {Fortney}, {Lian}, {Marley},
  {Freedman}, {Knutson}  \& {Charbonneau}}{{Showman}
  et~al.}{2009}]{showman2009}
{Showman} A.~P.,  {Fortney} J.~J.,  {Lian} Y.,  {Marley} M.~S.,  {Freedman}
  R.~S.,  {Knutson} H.~A.,   {Charbonneau} D.,  2009, \mn@doi [\apj]
  {10.1088/0004-637X/699/1/564}, \href
  {http://adsabs.harvard.edu/abs/2009ApJ...699..564S} {699, 564}

\bibitem[\protect\citeauthoryear{{Snellen}, {de Kok}, {de Mooij}  \&
  {Albrecht}}{{Snellen} et~al.}{2010}]{snellen2010}
{Snellen} I.~A.~G.,  {de Kok} R.~J.,  {de Mooij} E.~J.~W.,   {Albrecht} S.,
  2010, \mn@doi [\nat] {10.1038/nature09111}, \href
  {http://adsabs.harvard.edu/abs/2010Natur.465.1049S} {465, 1049}

\bibitem[\protect\citeauthoryear{{Spiegel}, {Silverio}  \& {Burrows}}{{Spiegel}
  et~al.}{2009}]{spiegel2009}
{Spiegel} D.~S.,  {Silverio} K.,   {Burrows} A.,  2009, \mn@doi [\apj]
  {10.1088/0004-637X/699/2/1487}, \href
  {http://adsabs.harvard.edu/abs/2009ApJ...699.1487S} {699, 1487}

\bibitem[\protect\citeauthoryear{{Steinrueck}, {Parmentier}, {Showman},
  {Lothringer}  \& {Lupu}}{{Steinrueck} et~al.}{2019}]{steinrueck2018}
{Steinrueck} M.~E.,  {Parmentier} V.,  {Showman} A.~P.,  {Lothringer} J.~D.,
  {Lupu} R.~E.,  2019, \mn@doi [\apj] {10.3847/1538-4357/ab2598}, \href
  {https://ui.adsabs.harvard.edu/abs/2019ApJ...880...14S} {880, 14}

\bibitem[\protect\citeauthoryear{{Stevenson} et~al.,}{{Stevenson}
  et~al.}{2014}]{stevenson2014}
{Stevenson} K.~B.,  et~al., 2014, \mn@doi [Science] {10.1126/science.1256758},
  \href {http://adsabs.harvard.edu/abs/2014Sci...346..838S} {346, 838}

\bibitem[\protect\citeauthoryear{{Stevenson} et~al.,}{{Stevenson}
  et~al.}{2017}]{stevenson2017}
{Stevenson} K.~B.,  et~al., 2017, \mn@doi [\aj] {10.3847/1538-3881/153/2/68},
  \href {http://adsabs.harvard.edu/abs/2017AJ....153...68S} {153, 68}

\bibitem[\protect\citeauthoryear{Sukoriansky, Dikovskaya  \&
  Galperin}{Sukoriansky et~al.}{2007}]{Sukoriansky2007}
Sukoriansky S.,  Dikovskaya N.,   Galperin B.,  2007, \mn@doi [Journal of the
  Atmospheric Sciences] {10.1175/jas4013.1}, 64, 3312

\bibitem[\protect\citeauthoryear{{Tan} \& {Komacek}}{{Tan} \&
  {Komacek}}{2019}]{tan2019}
{Tan} X.,  {Komacek} T.~D.,  2019, \mn@doi [\apj] {10.3847/1538-4357/ab4a76},
  \href {https://ui.adsabs.harvard.edu/abs/2019ApJ...886...26T} {886, 26}

\bibitem[\protect\citeauthoryear{{Tan} \& {Showman}}{{Tan} \&
  {Showman}}{2020}]{tan2020}
{Tan} X.,  {Showman} A.~P.,  2020, arXiv e-prints, \href
  {https://ui.adsabs.harvard.edu/abs/2020arXiv200512152T} {p. arXiv:2005.12152}

\bibitem[\protect\citeauthoryear{{Tennyson} et~al.,}{{Tennyson}
  et~al.}{2016}]{tennyson2016}
{Tennyson} J.,  et~al., 2016, \mn@doi [Journal of Molecular Spectroscopy]
  {10.1016/j.jms.2016.05.002}, \href
  {https://ui.adsabs.harvard.edu/abs/2016JMoSp.327...73T} {327, 73}

\bibitem[\protect\citeauthoryear{{Tremblin} et~al.,}{{Tremblin}
  et~al.}{2017}]{tremblin2017}
{Tremblin} P.,  et~al., 2017, \mn@doi [\apj] {10.3847/1538-4357/aa6e57}, \href
  {https://ui.adsabs.harvard.edu/abs/2017ApJ...841...30T} {841, 30}

\bibitem[\protect\citeauthoryear{{Visscher}, {Lodders}  \& {Fegley}}{{Visscher}
  et~al.}{2010}]{visscher2010}
{Visscher} C.,  {Lodders} K.,   {Fegley} Bruce J.,  2010, \mn@doi [\apj]
  {10.1088/0004-637X/716/2/1060}, \href
  {https://ui.adsabs.harvard.edu/abs/2010ApJ...716.1060V} {716, 1060}

\bibitem[\protect\citeauthoryear{{West} et~al.,}{{West}
  et~al.}{2016}]{west2016}
{West} R.~G.,  et~al., 2016, \mn@doi [\aap] {10.1051/0004-6361/201527276},
  \href {https://ui.adsabs.harvard.edu/abs/2016A&A...585A.126W} {585, A126}

\bibitem[\protect\citeauthoryear{{Wong} et~al.,}{{Wong}
  et~al.}{2015}]{wong2015_wasp14}
{Wong} I.,  et~al., 2015, \mn@doi [\apj] {10.1088/0004-637X/811/2/122}, \href
  {https://ui.adsabs.harvard.edu/abs/2015ApJ...811..122W} {811, 122}

\bibitem[\protect\citeauthoryear{{Wong} et~al.,}{{Wong}
  et~al.}{2016}]{wong2016}
{Wong} I.,  et~al., 2016, \mn@doi [\apj] {10.3847/0004-637X/823/2/122}, \href
  {https://ui.adsabs.harvard.edu/abs/2016ApJ...823..122W} {823, 122}

\bibitem[\protect\citeauthoryear{{Wordsworth}}{{Wordsworth}}{2015}]{wordsworth2015}
{Wordsworth} R.,  2015, \mn@doi [\apj] {10.1088/0004-637X/806/2/180}, \href
  {https://ui.adsabs.harvard.edu/abs/2015ApJ...806..180W} {806, 180}

\bibitem[\protect\citeauthoryear{Yamazaki, Skeet  \& Read}{Yamazaki
  et~al.}{2004}]{YAMAZAKI2004423}
Yamazaki Y.,  Skeet D.,   Read P.,  2004, \mn@doi [Planetary and Space Science]
  {https://doi.org/10.1016/j.pss.2003.06.006}, 52, 423

\bibitem[\protect\citeauthoryear{{Zahn}}{{Zahn}}{1992}]{1992A&A...265..115Z}
{Zahn} J.-P.,  1992, \aap, \href
  {http://adsabs.harvard.edu/abs/1992A%26A...265..115Z} {265, 115}

\bibitem[\protect\citeauthoryear{{Zellem} et~al.,}{{Zellem}
  et~al.}{2014}]{zellem2014}
{Zellem} R.~T.,  et~al., 2014, \mn@doi [\apj] {10.1088/0004-637X/790/1/53},
  \href {https://ui.adsabs.harvard.edu/abs/2014ApJ...790...53Z} {790, 53}

\bibitem[\protect\citeauthoryear{{Zhang} \& {Showman}}{{Zhang} \&
  {Showman}}{2017}]{zhang2017}
{Zhang} X.,  {Showman} A.~P.,  2017, \mn@doi [\apj]
  {10.3847/1538-4357/836/1/73}, \href
  {https://ui.adsabs.harvard.edu/abs/2017ApJ...836...73Z} {836, 73}

\bibitem[\protect\citeauthoryear{{Zhang} et~al.,}{{Zhang}
  et~al.}{2018}]{zhang2018}
{Zhang} M.,  et~al., 2018, \mn@doi [\aj] {10.3847/1538-3881/aaa458}, \href
  {https://ui.adsabs.harvard.edu/abs/2018AJ....155...83Z} {155, 83}

\makeatother
\end{thebibliography}



\appendix

\section{2D Flow}
\label{sec:2D}

While flow along isochors is what is seen in the atmospheres of Jupiter and Saturn~\citep{doi:10.1002/2017GL072628}, and is similar to the shallow atmosphere assumption used by many General Circulation Models~\citep[see e.g.][]{YAMAZAKI2004423}, it is worth examining the justification for this approximation.
Consider the equation of mass conservation,
\begin{align}
	\frac{\partial \boldsymbol{u}}{\partial t} + \nabla\cdot(\rho \boldsymbol{u}) = 0,
    \label{eq:cont0}
\end{align}
where $\rho$ is the density and $\boldsymbol{u}$ is the flow velocity in the frame corotating with the planet, which has mean angular velocity $\boldsymbol{\Omega}$.
In steady state the first term vanishes and we find
\begin{align}
	\nabla\cdot(\rho \boldsymbol{u}) = 0.
    \label{eq:steadycont}
\end{align}
Expanding equation~\eqref{eq:steadycont} in spherical coordinates we obtain
\begin{align}
	0 &= u_r\left(2 - \frac{r}{h_{\rho}} + r\frac{\partial \ln u_r}{\partial r}\right)\nonumber\\
    &+ u_\theta\left(\frac{\partial\ln\rho}{\partial\theta} + \frac{\partial\ln u_\theta}{\partial \theta} + \cot(\theta)\right)\nonumber\\
    &+ u_\phi \left(\frac{\partial \ln u_\phi}{\partial \phi}\right) + \frac{1}{\sin\theta}\frac{\partial \ln \rho}{\partial \phi},
    \label{eq:cont}    
\end{align}
where $\theta$ is the colatitude, $\phi$ is the azimuthal coordinate, $r$ is the spherical radial coordinate and $h_\rho$ is the density scale height
\begin{align}
	h_{\rho} \equiv -\frac{dr}{d\ln \rho}.
\end{align}
From dimensional analysis we expect each derivative to be set by some combination of the relevant length-scales in the system.
For derivatives in $\theta$ and $\phi$, barring large dimensionless numbers in the system, the relevant scale is just $\pi$, the angular separation between antipodes.
For derivatives in $r$ there are two potential scales: $r$ and $h_\rho$.
Hence  we expect these derivatives to produce factors between the orders of $r^{-1}$ and $h_\rho^{-1}$.
From this we see that the factor multiplying $u_r$ in equation~\eqref{eq:cont} is of order $r/h_\rho$ while those of $u_\theta$ and $u_\phi$ are of order unity as is the term independent of $\boldsymbol{u}$.
We therefore expect
\begin{align}
	|u_r| \approx \frac{h_{\rho}}{r} \left(|u_\theta| + |u_\phi|\right).
	\label{eq:cont2}
\end{align}
Because $h_\rho \ll r$,
\begin{align}
	|u_r| \ll \max(|u_\theta|,|u_\phi|),
\end{align}
and so the flow is strongly constrained to run in the angular directions.
When the density gradient is radial this is equivalent to saying that it runs along isochors.

When the density gradient is not radial a similar argument holds, but the relevant velocity component is that along $\nabla \rho$ and the relevant scale is set by $|\nabla \ln \rho|$.
So long as this remains large relative to derivatives in the other directions\footnote{We expect this to be the case so long as the angle between the density and pressure gradients is small.} the argument generalises straightforwardly.

Finally note that because the scale height is so much smaller than the radius the flow remains constrained to run along isochors even if there are large dimensionless numbers which serve to increase the scale of the angular derivatives, as in the case of rapidly rotating systems.

\section{Asymptotic Expansion}
\label{appen:cases}

We are interested in approximating the roots of equations of the form
\begin{align}
	\sum_{n=1}^N a_n x^{p_n} = 1
    \label{eq:sum}
\end{align}
where $N \geq 1$, $a_n \geq 0$ for all $n$, $p_j > p_i > 0$ for all $i > j$ and at least one $a_n > 0$.
In particular we are interested in producing asymptotic expansions of the unique positive real solution in the limit where one $a_n$ becomes much larger than all others.
In this limit the solution is just
\begin{align}
	x_n \approx a_n^{-1/p_n}.
	\label{eq:sol}
\end{align}
This holds so long as $a_n x_n^{p_n} \gg a_m x_n^{p_m}$ for all $m \neq n$.
Extending this down to the marginal case we require that
\begin{align}
a_n x_n^{p_n} \geq a_m x_n^{p_m}
\end{align}
for all $m \neq n$.
Using equation~\eqref{eq:sol}, this may be written as
\begin{align}
1 \geq a_m a_n^{-p_m/p_n},
\end{align}
or
\begin{align}
a_n^{p_m} \geq a_m^{p_n}
\label{eq:cond}
\end{align}
for all $m \neq n$.

A requirement to use equation~\eqref{eq:sol} is that inequality~\eqref{eq:cond} must be solved by at least one choice of $n$.
More valid choices are permitted only when equality is achieved.
To see this suppose that both $n$ and $n'$ are valid choices.
Then by two applications of inequality~\eqref{eq:cond}
\begin{align}
	a_n^{p_{n'}} \geq a_{n'}^{p_n} \geq a_n^{p_{n'}},
\end{align}
from which it follows that $a_n^{p_{n'}} = a_{n'}^{p_n}$.
Hence in such cases
\begin{align}
	x_{n'} = a_{n'}^{-1/p_{n'}} = a_n^{-1/p_n} = x_n,
\end{align}
so for the purposes of determining the solution for $x$ the existence of multiple valid choices does not matter.
Note that this also means that, as a function of the coefficients $a_i$, the solution $x$ is continuous even when the valid choice for $n$ changes.

We now turn to proving that there is always at least one valid choice via induction.
In the case where $N=1$ the inequality is trivially satisfied because there are no $m \neq n$.
Next suppose that there is a solution $x_n$ for some $N$ and we modify the equation to contain the term $a_{N+1} x^{p_{N+1}}$.
Then by assumption
\begin{align}
	a_{m < N+1}^{p_n} \leq a_{n}^{p_{m < N+1}}.
\end{align}
If $a_{N+1}^{p_n} \leq a_{n}^{p_{N+1}}$ then we are done because the new term is compatible with the same solution and does not support a different solution.
Otherwise suppose that the solution is now given by $x_{N+1} = a_{N+1}^{-1/p_{N+1}}$.
This requires that
\begin{align}
	a_{m}^{p_{N+1}} \leq a_{N+1}^{p_m}.
\end{align}
for all $m < N+1$.
By assumption, however, we have $a_{N+1}^{p_n} \geq a_{n}^{p_{N+1}}$, so
\begin{align}
	a_{N+1}^{p_m} \geq a_{n}^{p_{N+1} p_{m} / p_{n}} \geq \left(a_m^{p_n}\right)^{p_{N+1}/p_n} = a_{m}^{p_{N+1}},
\end{align}
where we have used the fact that $a_n^{p_m} \geq a_m^{p_n}$ for all $m < N$.
It follows that the new solution is indeed valid.
Hence by induction we see that for all $N \geq 1$ there is a choice of $n$ satisfying inequality~\eqref{eq:cond}.

Because we have extended the asymptotic solutions down to the marginal limit, in which different terms in equation~\eqref{eq:sum} contribute comparably, it is important to consider the error incurred in this limit.
To bound this error note that it is worst when every term in equation~\eqref{eq:sum} contributes equally, which occurs when inequality~\eqref{eq:cond} is saturated.
In this case the error in equation~\eqref{eq:sum} is a factor of $N$.
We do not, however, care particularly about the error in equation~\eqref{eq:sum}.
Rather we are interested in the error of our approximation of its solution.
As a crude estimate therefore we must reduce the contribution of the average term in equation~\eqref{eq:sum} by a factor of $N$.
The resulting solution is reduced by a factor of order $N^{-1/p}$, where $p$ is an appropriately weighted average of the exponents in equation~\eqref{eq:sum}.
Hence the error is bounded by a factor of order unity when $N$ is small and $p$ is of order unity, but could become large if some $p_i$ are small or if $N$ is large.
This does not occur in the cases of interest, and even if it were to occur the worst-case scenario in which inequality~\eqref{eq:cond} saturates represents a very small portion of parameter space, so it is relatively safe to use these asymptotic solutions.

We begin with the convective case.
As we showed in Section~\ref{sec:summary}, $\lambda \ll 1$ so equation~\eqref{eq:conv_non_dim_0} becomes
\begin{align}
	\frac{1}{3}\bar{u}\left(\bar{u} + \bar{\varv}_c \right) = 1.
\end{align}
This allows two cases, with $a_1 = \bar{\varv}_c/3$, $p_1 = 1$, $a_2 = 1/3$ and $p_2 = 2$.

Next we turn to equation~\eqref{eq:rad_non_dim_0}.
This is not immediately in the form of equation~\eqref{eq:sum}, but we may divide the parameter space into two regions such that it is of the appropriate form in each.
In the first regime we take $\bar{u} > \bar{\eta}$, so that equation~\eqref{eq:rad_non_dim_0} may be approximated as
\begin{align}
	\bar{u}^2 \left(\lambda + C\right) = 1.
\end{align}
Recalling that $\lambda \ll 1$ this reduces to
\begin{align}
	\bar{u} = C^{-1/2},
    \label{eq:uc}
\end{align}
with the condition $C \bar{\eta}^2 < 1$.

The remaining region in parameter space has $C \bar{\eta}^2 > 1$.
In this region equation~\eqref{eq:rad_non_dim_0} may be approximated as
\begin{align}
	\bar{u}^2 \left(\lambda + C\frac{\bar{u}}{\bar{\eta}}\right) = 1.
\end{align}
In order to put this in the form of equation~\eqref{eq:sum} we further approximate equation~\eqref{eq:unsimp} by
\begin{align}
	\lambda \approx \lambda_0\left(1 + \sqrt{\frac{\bar{u}_{\Omega}}{\bar{u}}}\right),
\end{align}
where we have used $\varv_c = 0$ in radiative zones.
This is a rather crude approximation but it preserves the scaling of $\lambda$ with each of $\bar{u}$ and $\bar{u}_\Omega$.
Hence we obtain
\begin{align}
	\bar{u}^2 \left(\lambda_0\left(1 + \sqrt{\frac{\bar{u}_\Omega}{\bar{u}}}\right) + C\frac{\bar{u}}{\bar{\eta}}\right) = 1.
\end{align}
This is now in the appropriate form, with $a_1 = \lambda_0\sqrt{\bar{u}_\Omega}$, $p_1 = 3/2$, $a_2 = \lambda_0$, $p_2 = 2$, $a_3 = C/\bar{\eta}$ and $p_3 = 3$.

All that remains is to show that these three cases derived from equation~\eqref{eq:sum} are consistent with equation~\eqref{eq:uc} and that these are continuous across the boundary $C \bar{\eta}^2 = 1$.
To show consistency consider the solution $x_3$ corresponding to $a_3$ and $p_3$.
This yields
\begin{align}
	\frac{x_3}{\bar{\eta}} = C^{-1/3} \bar{\eta}^{-2/3} = (C \bar{\eta}^2)^{-1/3}.
\end{align}
When $C \bar{\eta}^2 > 1$ this is less than one.
Next consider a different solution $x_i$ corresponding to $a_i$ and $p_i$.
When this solution dominates, inequality~\eqref{eq:cond} yields
\begin{align}
		a_i^{3} \geq a_3^{p_i},
\end{align}
from which it follows that
\begin{align}
	x_i = a_i^{-1/p_i} \leq a_3^{-1/3} = x_3.
\end{align}
So when $x_i$ is the solution it is smaller than $x_3$, and $x_3$ is always less than $\bar{\eta}$ when $C \bar{\eta}^2 > 1$.
Hence $\bar{u} > \bar{\eta}$ occurs if and only if $C\bar{\eta}^2 < 1$.
It follows that the solutions are consistent.

Next to show that the solutions are continuous across the boundary $C\bar{\eta}^2 = 1$, note that equation~\eqref{eq:uc} gives $\bar{u} = 1/\sqrt{C}$.
On this boundary $x_3 = 1/\sqrt{C}$, so this pair is continuous.
We now claim that $x_3$ is the only possible solution from among $\{x_i\}$ on the boundary.
This is because when $C \bar{\eta}^2 = 1$ inequality~\eqref{eq:cond} applied for $n=3$ yields
\begin{align}
	C &\geq \lambda_0\\
\intertext{and}
	C &\geq \left(\lambda_0 \bar{u}_\Omega^{1/2}\right)^{4/3}.
\end{align}
The former is straightforwardly satisfied because $C \approx 10^{-1}$ while $\lambda_0$ is orders of magnitude smaller.
Noting that
\begin{align}
\lambda_0 \bar{u}_\Omega^{1/2} = \lambda \bar{u}^{1/2},
\end{align}
we see that the latter condition may be written as
\begin{align}
	C \geq \left(\lambda \bar{u}^{1/2}\right)^{4/3}.
\end{align}
With $\bar{u} = x_3 = C^{-1/2}$ the condition reduces to $C \geq \lambda$.
In Section~\ref{sec:summary} we showed that $\lambda$ was several orders of magnitude smaller than unity while $C \approx 10^{-1}$, so this is satisfied.
Hence $x_3$ dominates along the boundary, so the solution is continuous across the boundary.

\section{Verifying Assumptions}
\label{appen:verify}

We now verify our earlier assumptions.
First, we have claimed that two-dimensional flow is preserved under all circumstances, such that $l_h \gg l_v$.
This is clearly true in the case where $l_h = r$ because $l_v \approx h \ll r$ so we focus on the case where $l_h = 2\pi \sqrt{r v_{\rm turb}/\Omega}$.
Setting $l_h = l_v$ defines a critical rotation rate
\begin{align}
	\Omega_0 = \frac{4\pi^2 r v_{\rm turb}}{h^2}
\end{align}
above which $l_h < l_v$.
Noting that $v_{\rm turb} > u$, we see that
\begin{align}
	\Omega_0 > 4\pi^2 \left(\frac{r}{h}\right)^2 \frac{u}{r}.
\end{align}
By comparison the breakup rotation rate is
\begin{align}
	\Omega_{\rm b} = \sqrt{\frac{g}{r}},
\end{align}
so
\begin{align}
	\frac{\Omega_0^2}{\Omega_{\rm b}^2} > 16\pi^4 \left(\frac{r}{h}\right)^3  \frac{u^2}{g h}.
\end{align}
Noting that $gh \approx c_{\rm s}^2$ we see that
\begin{align}
	\frac{\Omega_0^2}{\Omega_{\rm b}^2} > 16\pi^4 \left(\frac{r}{h}\right)^3  \frac{u^2}{c_{\rm s}^2}.
\end{align}
For the planets and regions of interest $h \approx 10^7{\rm cm}$, $r \approx 10^{10}{\rm cm}$, $u \approx 10^{4}{\rm cm\,s^{-1}}$ and $c_{\rm s} \approx 10^5{\rm cm\,s^{-1}}$.
It follows that this ratio is of order $10^{10}$, meaning that even for planets with much weaker flows rotating near breakup the horizontal scale of motion remains larger than the vertical one.

Secondly, we have claimed that $u^2 \boldsymbol{u}\cdot \nabla \ln \rho(\boldsymbol{r})$ may be dropped from equation~\eqref{eq:balance4}.
This is equivalent to dropping $u^2 \nabla \cdot \boldsymbol{u}$ from equation~\eqref{eq:balance3}.
Using the same length-scales as before we estimate
\begin{align}
|\nabla\cdot\boldsymbol{u}| \approx \frac{u_r}{l_v} + \frac{u}{l_h},
\label{eq:divuu}
\end{align}
where $u_r$ is the radial (vertical) component of the motion.
We next insert equation~\eqref{eq:cont2} into equation~\eqref{eq:divuu}.
Importantly though $r^{-1}$ in the former represents an angular derivative and so should be replaced with $l_v^{-1}$ in the latter.
Making this replacement and insertion and identifying $u \approx |u_\theta| + |u_\phi|$ we find
\begin{align}
|\nabla\cdot\boldsymbol{u}| \approx 2\frac{u}{l_h},
\end{align}
Hence
\begin{align}
u^2 \nabla \cdot \boldsymbol{u} \approx \frac{2 u^3}{l_h}.
\end{align}
By contrast the remaining terms in equation~\eqref{eq:balance3}, when balanced against one another, have magnitude $u_0^3/h$.
From Table~\ref{tab:cases} we see that $u < u_0$, and we know that $l_h \gg l_v$, so this term may indeed be dropped.

\section{Modifying the Radiative Transfer and Radiative-Convective Equilibrium Equations}\label{sec:rt_rceqm}

\begin{table*}
\caption{Summary of the radiative-convective (thermal) equilibrium conditions and the radiative transfer equations with their respective boundary conditions. We adopt notation, definitions and units consistent with \citet{gandhi2017} for all quantities for clarity. All physical quantities are separately calculated for the day and night side, except for the wind flux $D$, which is identical between the day and night.}
\renewcommand{\arraystretch}{2}
\begin{tabular}{|l|l|l|l|}\label{tab:eqn_summary_genesis}
           &  & \bf{Day} & \bf{Night} \\
\hline
\multirow{3}{*}{\bf{Radiative Transfer}} &       & $\left.\frac{\partial (f_\nu J_\nu)}{\partial \tau_\nu}\right|_{\tau = 0} = g_\nu J_\nu(0) - H_\mathrm{ext}$ & $\left.\frac{\partial (f_\nu J_\nu)}{\partial \tau_\nu}\right|_{\tau = 0} = g_\nu J_\nu(0) $  \\
                                    &                            & $\frac{\partial^2 (f_\nu J_\nu)}{\partial \tau_{\nu}^2} = J_\nu - S_\nu = \frac{\kappa_\nu (J_\nu-B_\nu)}{\kappa_\nu +\sigma_\nu}$ & $\frac{\partial^2 (f_\nu J_\nu)}{\partial \tau_{\nu}^2} = J_\nu - S_\nu = \frac{\kappa_\nu (J_\nu-B_\nu)}{\kappa_\nu +\sigma_\nu}$   \\
                                    &    & $\left.\frac{\partial (f_\nu J_\nu)}{\partial \tau_\nu}\right|_{\tau = \tau_\mathrm{max}} = \frac{1}{2}(B_\nu - J_\nu) +\frac{1}{3}\frac{\partial B_\nu}{\partial \tau_\nu}$   & $\left.\frac{\partial (f_\nu J_\nu)}{\partial \tau_\nu}\right|_{\tau = \tau_\mathrm{max}} = \frac{1}{2}(B_\nu - J_\nu) +\frac{1}{3}\frac{\partial B_\nu}{\partial \tau_\nu}$\\
\hline
\multirow{2}{*}{\bf{RC Equilibrium}} & & $\int_0^\infty \kappa_\nu (J_\nu-B_\nu) d\nu + \frac{\rho g}{4\pi}\frac{dF_\mathrm{conv}}{dP} = -D$ & $\int_0^\infty \kappa_\nu (J_\nu-B_\nu) d\nu + \frac{\rho g}{4\pi}\frac{dF_\mathrm{conv}}{dP} = D$ \\
 &                                              & $\int_0^\infty \frac{d(f_\nu J_\nu)}{d\tau_\nu} d\nu +\frac{F_\mathrm{conv}}{4\pi} = \frac{\sigma_R}{4 \pi}T_\mathrm{int}^4 - \int_{P_\mathrm{max}}^{P} \frac{D}{\rho g} dP'$ & $\int_0^\infty \frac{d(f_\nu J_\nu)}{d\tau_\nu} d\nu +\frac{F_\mathrm{conv}}{4\pi} = \frac{\sigma_R}{4 \pi}T_\mathrm{int}^4 + \int_{P_\mathrm{max}}^{P} \frac{D}{\rho g} dP'$  \\
\hline
\end{tabular}
\end{table*}

\subsection{Radiative Transfer}\label{sec: rad transfer}

The radiative transfer is solved using the Feautrier method for both sides of the model planet. The conditions on the radiative transfer solution are summarised in Table~\ref{tab:eqn_summary_genesis}. We adopt notation that is identical to \citet{gandhi2017} for all quantities and units for consistency. 

The radiative transfer equation for both the day and night sides of our model atmosphere is given by 
\begin{align}
\frac{\partial^2 (f_\nu J_\nu)}{\partial \tau_{\nu}^2} = J_\nu - S_\nu = \frac{\kappa_\nu (J_\nu-B_\nu)}{\kappa_\nu +\sigma_\nu},
\end{align}
where $\tau_\nu$ refers to the frequency dependent optical depth, $\kappa_\nu$ is the absorption coefficient, $\sigma_\nu$ is the scattering coefficient and $S_\nu$ refers to the source function of radiation,
\begin{align}
S_\nu & = \frac{\kappa_\nu B_\nu + \sigma_\nu J_\nu}{\kappa_\nu + \sigma_\nu}.
\end{align}
Here, $B_\nu$ refers to the Planck function,
\begin{align}
B(T,\nu) & = \frac{2h \nu^3}{c^2}\frac{1}{e^\frac{h \nu}{k_b T} - 1}.
\end{align}
The first moment of the spectral radiance $J_\nu$ is defined in terms of the spectral radiance as
\begin{align}
J_\nu \equiv \frac{1}{2}\int_{-1}^{1} \mu I_\nu d\mu.
\end{align}
We also define the second and third moments $H_\nu$ and $K_\nu$ as
\begin{align}
H_\nu & \equiv \frac{1}{2}\int_{-1}^{1} \mu I(\mu)d\mu,\\
K_\nu & \equiv \frac{1}{2}\int_{-1}^{1} \mu^2 I(\mu)d\mu,
\end{align}
and the quantity $f$ as
\begin{align}
f_\nu & \equiv K_\nu/J_\nu = \frac{\int_{-1}^{1}I(\mu)\mu^2 d\mu}{\int_{-1}^{1}I(\mu)d\mu},\label{eqn:f_nu}
\end{align}

\subsubsection{Top Boundary Conditions}
The top boundary conditions do differ, given the night side profile has no incident stellar irradiation incident upon it. Therefore the top boundary condition for radiative transfer is,
\begin{align}
\left.\frac{\partial (f_\nu^n J_\nu^n)}{\partial \tau_\nu^n}\right|_{\tau = 0} = g_\nu^n J_\nu^n(0), \label{eqn:topbc_n}
\end{align}
where the superscript $n$ refers to all quantities calculated for the night side and $g_\nu$ is defined as
\begin{align}
g_{\nu} & \equiv \frac{H_{\nu}(\tau = 0)}{J_{\nu}(\tau = 0)} = \frac{\int_{0}^{1}I(\mu,\tau =0)\mu d\mu}{\int_{-1}^{1}I(\mu,\tau =0)d\mu}.
\end{align}
The day side of our model atmosphere has the full undiluted irradiation,
\begin{align}
\left.\frac{\partial (f_\nu^d J_\nu^d)}{\partial \tau_\nu^d}\right|_{\tau = 0} = g_\nu^d J_\nu^d(0) - H_\mathrm{ext}, \label{eqn:topbc},
\end{align}
where the superscript $d$ refers to quantities on the day side. 

\subsubsection{Bottom Boundary Condition}
The bottom boundary condition is the diffusion approximation and remains unchanged for both the day and night side,
\begin{align}
\left.\frac{\partial (f_\nu^n J_\nu^n)}{\partial \tau_\nu^n}\right|_{\tau = \tau_\mathrm{max}} &= \frac{1}{2}(B_\nu^n - J_\nu^n) +\frac{1}{3}\frac{\partial B_\nu^n}{\partial \tau_\nu^n}.\label{eqn:bottombc_n}, \\
\left.\frac{\partial (f_\nu^d J_\nu^d)}{\partial \tau_\nu^d}\right|_{\tau = \tau_\mathrm{max}} &= \frac{1}{2}(B_\nu^d - J_\nu^d) +\frac{1}{3}\frac{\partial B_\nu^d}{\partial \tau_\nu^d}.\label{eqn:bottombc}
\end{align}

\subsection{Modifying the Equilibrium Equations}\label{sec:rad eqm}

The sources of energy for the day side in our model are the internal heat flux, the external stellar irradiation and the wind flux. In every layer of the atmosphere we must conserve the total energy flowing into it with the energy flowing out. The atmosphere is computed in radiative-convective equilibrium using the complete linearisation (Rybicki) method. On the day side, we have an external flux from the star, an internal flux emanating from the planet interior, and a wind that removes energy out of each layer of the atmosphere. On the night side, the same internal heat is transported up, but now with the wind flux being additive energy into each layer of the atmosphere. We assume that the energy leaving the day side at a pressure $P$ is deposited at the same pressure on the night side, and thus that the wind flows along isobars. We need to modify the radiative-convective equilibrium conditions of the day and night sides accordingly in order to take the wind flux $D$ into account.

There are two forms of the radiative-convective equilibrium equations which are used in the upper and lower parts of the atmosphere. These equations are mathematically identical but numerically have different behaviour \citep{hubenybook}. We use the integral form when the optical depth $\tau << 1$ and the differential form in the deep atmosphere. The conditions without any wind are
\begin{align}\label{eqn:reqmint}
& \int_0^\infty \kappa_\nu (J_\nu-B_\nu) d\nu + \frac{\rho g}{4\pi}\frac{dF_\mathrm{conv}}{dP} = 0,\\ 
& \int_0^\infty \frac{d(f_\nu J_\nu)}{d\tau_\nu} d\nu +\frac{F_\mathrm{conv}}{4\pi} = \frac{\sigma_R}{4 \pi}T_\mathrm{int}^4,
\label{eqn:reqmdiff}
\end{align}
where $\sigma_R$ is the Stefan-Boltzmann constant and $F_\mathrm{conv}$ refers to the convective flux, only applied in regions where the temperature gradient exceeds the adiabatic gradient, $\nabla > \nabla_\mathrm{ad}$. We use mixing length theory \citep{kippenhahn2012} to derive the convective flux in such regions of the atmosphere. Further detail on these conditions and their derivations can be found in \citet{gandhi2017} and \citet{hubeny2017}.

To modify these equations to include the flux lost/gained from the day/night sides we consider the method in \cite{burrows2008} to include a source and sink of energy in the radiative-convective equilibrium conditions. Firstly we describe the day side, where the wind removes flux from each layer of the atmosphere. When a pressure dependent wind flux $D$ is applied the equations for the day side atmosphere are given by
\begin{align}\label{eqn:reqmint_wind}
& \int_0^\infty \kappa_\nu^d (J_\nu^d-B_\nu^d) d\nu + \frac{\rho^d g}{4\pi}\frac{dF_\mathrm{conv}^d}{dP} = -D,\\ 
& \int_0^\infty \frac{d(f_\nu^d J_\nu^d)}{d\tau_\nu^d} d\nu +\frac{F_\mathrm{conv}^d}{4\pi} = \frac{\sigma_R}{4 \pi}T_\mathrm{int}^4 - \int_{P_\mathrm{0}}^{P} \frac{D}{\rho^d g} dP'.
\label{eqn:reqmdiff_wind}
\end{align}
As before, the superscript $d$ refers to the day side. The night side now has the corresponding flux $D$ added to it, and therefore these conditions are given by
\begin{align}\label{eqn:reqmint_wind_n}
& \int_0^\infty \kappa_\nu (J_\nu^n-B_\nu^n) d\nu + \frac{\rho^n g}{4\pi}\frac{dF_\mathrm{conv}^n}{dP} = D,\\ 
& \int_0^\infty \frac{d(f_\nu^n J_\nu^n)}{d\tau_\nu^n} d\nu +\frac{F_\mathrm{conv}^n}{4\pi} = \frac{\sigma_R}{4 \pi}T_\mathrm{int}^4 + \int_{P_\mathrm{0}}^{P} \frac{D}{\rho^n g} dP'.
\label{eqn:reqmdiff_wind_n}
\end{align}
We adopt both the differential form and the integral form for the equilibrium conditions and have a switch over point in the atmosphere near $\tau =1$. This is because the integral and differential form (equations \ref{eqn:reqmint} and \ref{eqn:reqmdiff}) are the most numerically stable at low and high optical depths respectively.


\bsp	
\label{lastpage}
\end{document}